\documentclass[aps,pra,superscriptaddress,notitlepage,floatfix,twocolumn,longbibliography,reprint]{revtex4-1}
\usepackage{amssymb, amsmath, amsfonts} 
\usepackage{graphicx}
\usepackage[hidelinks,hypertexnames=false]{hyperref}
\usepackage[all]{hypcap}
\usepackage{outlines}
\usepackage{enumitem}
\usepackage{bm}
\usepackage{braket}

\renewcommand{\Re}{\operatorname{Re}}
\renewcommand{\Im}{\operatorname{Im}}

\newcommand{\Dpc}{\Delta}
\newcommand{\nbar}{\bar{n}}
\newcommand{\cavityW}{ {\omega_\text{c}} }
\newcommand{\probeW}{ {\omega_\text{p}} }

\newcommand{\Nosc}{N}
\newcommand{\Nsamp}{ {n_s} }
\newcommand{\Ntime}{ {N_t} }
\newcommand{\Psn}{ {P_\text{SN}} }

\newcommand{\cavHat}{\hat{c}}
\newcommand{\cavDag}{ {\cavHat^\dagger} }
\newcommand{\cavPM}{ {\cavHat^\text{PM}} }
\newcommand{\cavAM}{ {\cavHat^\text{AM}} }
\newcommand{\out}{ {\cavHat_\text{out}} }

\newcommand{\ahat}{\hat{a}}
\newcommand{\adag}{\ahat^\dagger}

\newcommand{\vac}{\hat{\xi}}
\newcommand{\vacdag}{ {\vac^\dagger} }
\newcommand{\vacPM} { {\vac^\text{PM}} }
\newcommand{\vacAM}{ {\vac^\text{AM}} }

\newcommand{\bath}{\hat{\eta}}
\newcommand{\bathdag}{ {\bath^\dagger} }
\newcommand{\bathPM}{ {\bath^\text{PM}} }
\newcommand{\bathAM}{ {\bath^\text{AM}} }
\newcommand{\bathTemp}{\nu}

\newcommand{\tf}{{t_f}}
\newcommand{\Tr}{\mathsf{T}}
\renewcommand{\vec}[1]{{\bm{#1}}}
\newcommand{\mat}[1]{\mathrm{#1}}
\newcommand{\est}[1]{\breve{#1}}

\newcommand{\Q}{ Q }
\newcommand{\rawVec}{ \est{ \vec{q} } }
\newcommand{\quadVec}{ \vec{\Q} }
\newcommand{\gFilt}{\gamma}
\newcommand{\gOpt}{ {\gFilt_\text{opt}} }
\newcommand{\sig}{ \hat{S} }

\newcommand{\matTh}{\mat{T}}
\newcommand{\matBA}{\mat{B}}
\newcommand{\matSN}{\mat{M}}

\newcommand{\sigDiff}{D}
\newcommand{\noiseQuad}{d}

\newcommand{\imp}{{\Delta n}}
\newcommand{\Copt}{{C_\text{opt}}}

\begin{document}

\title{Simultaneous retrodiction of multi-mode optomechanical systems using matched filters}

\author{Jonathan Kohler}
\email{jkohler@berkeley.edu}
\affiliation{Department of Physics, University of California, Berkeley, California 94720, USA}
\author{Justin A.\ Gerber}
\affiliation{Department of Physics, University of California, Berkeley, California 94720, USA}
\author{Emma Deist}
\affiliation{Department of Physics, University of California, Berkeley, California 94720, USA}
\author{Dan M.\ Stamper-Kurn}
\email{dmsk@berkeley.edu}
\affiliation{Department of Physics, University of California, Berkeley, California 94720, USA}
\affiliation{Materials Sciences Division, Lawrence Berkeley National Laboratory, Berkeley, California  94720, USA}

\begin{abstract}
Generation and manipulation of many-body entangled states is of considerable interest, for applications in quantum simulation or sensing, for example.
Measurement and verification of the resulting many-body state presents a formidable challenge, however, which can be simplified by multiplexed readout using shared measurement resources.
In this work, we analyze and demonstrate state retrodiction for a system of optomechanical oscillators coupled to a single-mode optical cavity.
Coupling to the shared cavity field facilitates simultaneous optical measurement of the oscillators' transient dynamics at distinct frequencies.
Optimal estimators for the oscillators' initial state can be defined as a set of linear matched filters, derived from a detailed model for the detected homodyne signal.
We find that the optimal state estimate for optomechanical retrodiction is obtained from high-cooperativity measurements, reaching estimate sensitivity at the Standard Quantum Limit (SQL).
Simultaneous estimation of the state of multiple oscillators places additional limits on the estimate precision, due to the diffusive noise each oscillator adds to the optomechanical signal.
However, we show that the sensitivity of simultaneous multi-mode state retrodiction reaches the SQL for sufficiently well-resolved oscillators.
Finally, an experimental demonstration of two-mode retrodiction is presented, which requires further accounting for technical fluctuations of the oscillator frequency.
\end{abstract}

\maketitle

Building many-body quantum systems by assembling ensembles of well-controlled quantum modes with tunable interactions is a promising path toward quantum simulation and quantum information processing.
The increased dimensionality of many-body systems, however, makes measurement of entangled states challenging, because of the large number of observables required to fully characterize the quantum state.
The physical resources necessary to perform these measurements can be reduced by using a shared measurement `bus,' coupled to multiple quantum degrees of freedom, facilitating multiplexed measurement of their quantum states, such as demonstrated with arrays of super-conducting qubits coupled to a common strip-line resonator~\cite{Jerger2012,Chen2012}.
Each mode can be independently measured by being sequentially coupled to the measurement bus, reading out each of their states with independent temporal modes of the output field~\cite{Palomaki2013}.
Alternatively, if the dynamics of each mode are spectrally resolved at distinct frequencies, the many-body state of a system can be simultaneously measured through a continuous weak measurement.

High-finesse optical resonators provide a particularly powerful tool for measuring and controlling the dynamics of diverse systems, demonstrated in cavity optomechanics~\cite{Aspelmeyer2014}, collective atomic spin optodynamics~\cite{Leroux2010,Vasilakis2015,Kohler2017}, and in the emerging field of cavity optomagnonics~\cite{Zhang2015a,Haigh2015,Osada2016, Liu2016, Kusminskiy2016}. 
Multiple modes of diverse systems can be simultaneously coupled to a common cavity field, facilitating simultaneous optical measurement~\cite{Botter2013}, in addition to long-range optically mediated interactions \cite{Shkarin2014,Spethmann2015,Kohler2018}.

The optical field leaking out of the cavity provides a continuous measurement of the system dynamics.
Retrodiction of past quantum states from continuous measurements has been theoretically described through back-propagation of an effect matrix~\cite{Gammelmark2013}, closely related to the quantum theory of smoothing~\cite{Tsang2009b}.
For Gaussian dynamics, these estimates allow a particularly compact description by evolving the phase space mean and covariance~\cite{Zhang2017,Huang2018,LammersThesis,Lammers2018}, recently demonstrated on an optomechanical system~\cite{Rossi2019}.
Related analysis employing linear filters applied to the output of an optical interferometer has been proposed for conditional quantum state preparation~\cite{Muller-Ebhardt2009} and verification~\cite{Miao2010}, applicable for tests of macroscopic quantum mechanics in gravitational-wave detectors such as LIGO.
For a multi-mode optomechanical system, estimation of a collective quadrature has also been proposed through temporal modulation of the measurement strength~\cite{Moore2016}.

\begin{figure}[bt!] 
\includegraphics[width=3.375in]{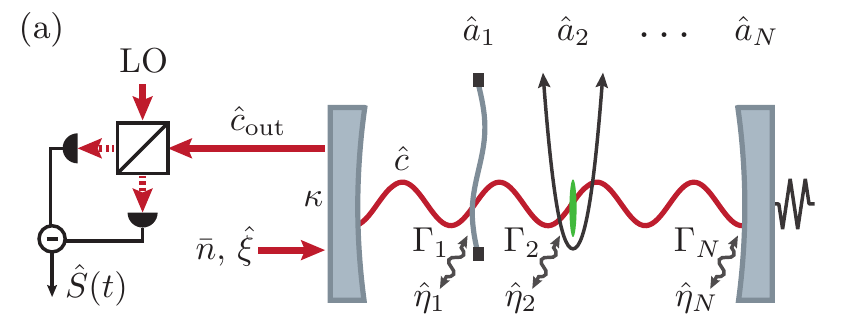}	
\includegraphics[width=3.375in]{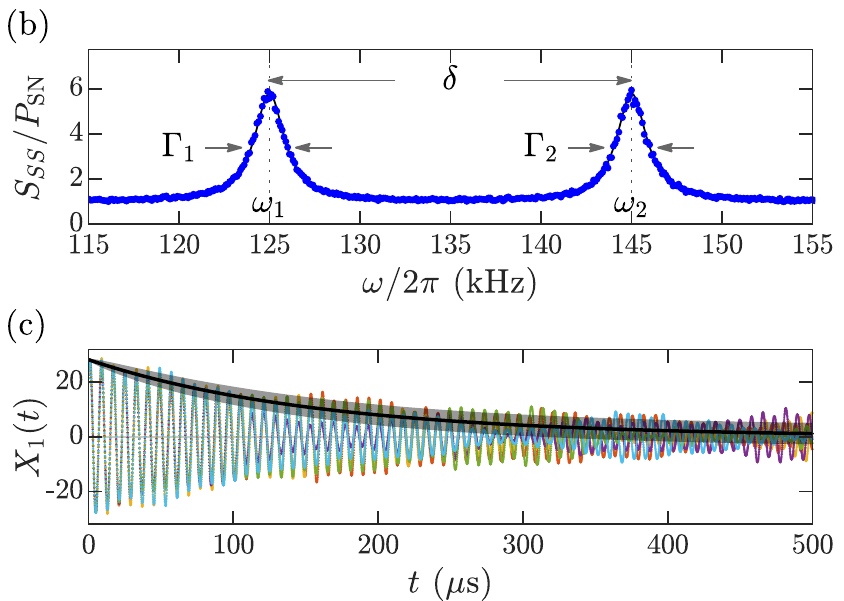}
\caption{
	(a) Schematic of a multi-mode optomechanical system, with a diverse ensemble of harmonic oscillators linearly coupled to a driven single-mode optical cavity.
	Simultaneous measurement of the motion of multiple oscillator modes can be performed through continuous homodyne measurement of the reflected optical phase quadrature.
	(b) Noise power-spectral density (PSD) for measurement of a two-mode optomechanical system, observed in the reflected homodyne phase quadrature for a resonantly driven cavity and normalized to the homodyne shot noise PSD.
	This spectrum reveals the stationary thermal and backaction-driven response of the mechanical oscillators, providing experimental calibration of the intrinsic oscillator frequencies $\omega_i$, damping rates $\Gamma_i$, and coupling strengths $g_i$.
	(c) Semi-classically simulated trajectories of a continuously measured oscillator, sampled from the same initial phase space coordinates, illustrating coherent decay of the initial state with an exponential envelope (black line) and increasing variance (shaded region) from accumulated diffusion driven by thermal noise and quantum backaction.
}
\label{fig:simulation}
\end{figure}

In this work, we consider estimation of the initial state, at time $t=0$, of a multi-mode optomechanical system of $\Nosc$ harmonic oscillators, illustrated in Fig.~\ref{fig:simulation}a, retrodicted using matched filters applied to continuous measurement of its \emph{subsequent} free evolution.
The oscillators are dispersively coupled to a common single-mode optical cavity, which is driven on resonance with a coherent probe.
The reflected optical field performs a continuous weak measurement of the sum of oscillator displacements, recorded using a balanced homodyne detector.
Knowledge of the coherent system evolution allows the initial state to be inferred from the observed transient dynamics.
However, measurement backaction, arising from quantum fluctuations of the cavity field, perturbs the trajectory of each oscillator's evolution, adding incoherent noise to the subsequent measurement record, which must be considered in obtaining the optimal state estimate.
We focus in particular on the experimentally relevant system of linear cavity optomechanics, allowing derivation of analytic results for the optimal state estimators which can be directly applied to experimental measurements.
However, the formalism developed in this work can be directly generalized for measurement and retrodiction of any linearizable system undergoing transient or non-steady state dynamics driven by Markovian noise.

Retrodiction can be illustrated by considering continuous measurement of the position $\hat{X}_1(t)$ of a single harmonic oscillator.
Estimates for the average position quadrature $\langle \hat{X}_1(0) \rangle$ and momentum quadrature $\langle \hat{P}_1(0) \rangle$ of the oscillator's initial state can be recovered from measurement of its subsequent coherent evolution 
\begin{align}
\langle \hat{X}_1(t) \rangle = e^{- \Gamma_1 t / 2} \Bigl( \langle \hat{X}_1(0) \rangle  \cos \omega_1 t + \langle \hat{P}_1(0) \rangle \sin \omega_1 t  \Bigr)
\text{,}
\end{align}
where $\omega_1$ and $\Gamma_1$ are the oscillators frequency and energy damping rate, respectively.
Incoherent noise from measurement backaction and the oscillator's intrinsic thermal bath also perturbs the oscillator's trajectory during measurement, as simulated in Fig.~\ref{fig:simulation}c.
The accumulated diffusion from these noise baths reduces the relative signal-to-noise in the measurement record at later times.

Quadrature estimators for the oscillator's initial state can be defined as linear filters of the recorded homodyne signal $\sig(t)$, which appropriately weight the measured signal at each subsequent time $t$.
The optimally `matched' filters must appropriately balance the coherent evolution and incoherent diffusion, in order to minimize the total estimate error.
The use of matched filters to recover signals of a known form from stationary additive Gaussian noise is well described in standard textbooks \cite{Wainstein1970} and commonly employed, for instance, for gravitational wave detection~\cite{LIGOAnalysis}.
However, for retrodiction of optomechanical systems, diffusion of the oscillators' state, driven by quantum backaction and the thermal bath, introduces non-stationary noise which accumulates throughout the observed transient signals, requiring careful accounting of the full two-time correlation of the signal noise.

For high-quality oscillators, where $\omega_i \gg \Gamma_i$, the initial state undergoes multiple coherent oscillations during the subsequent ringdown.
Retrodiction from the observed trajectory, therefore, obtains approximately equal information about each initial quadrature amplitude, recovering an estimate of the oscillator's initial state that is independent of the oscillator's phase.
The Heisenberg uncertainty principle
$ \langle \Delta \hat{X}_i^2 \rangle \langle \Delta \hat{P}_i^2 \rangle \ge 1/4 $
establishes a fundamental bound, known as the Standard Quantum Limit (SQL)~\cite{Caves1980a,BKT_QuantumNoise_1992}, for the minimum noise added by such phase-independent measurements
\begin{align}
\langle \Delta \hat{X}_i^2 \rangle = \langle \Delta \hat{P}_i^2 \rangle  = \imp_i \ge \frac{1}{2}
\text{,}
\end{align}
quantified here as an added effective thermal phonon occupation $\imp_i$.

In this work, we consider retrodiction of Gaussian states and demonstrate inference of two-mode squeezed states in a multi-mode optomechanical system.
Full tomography of a general many-body state involves estimation of the entire density matrix, which contains further information about all higher-order moments of the oscillator quadratures.
The following analysis could be extended to estimate these higher-order moments of the multi-mode state.
Though beyond the scope of this work, it would be worth considering which features of multi-mode quantum states can be retrodicted from such phase-independent optomechanical measurements~\cite{Nha2010} or how measurements beyond the SQL~\cite{Miao2010,Lei2016,Ockeloen-Korppi2016,Moller2017} can be performed simultaneously on multi-mode systems.

\subsection{Summary of main results}



We summarize here the primary conclusions of this work and give an overview of the following sections.
In Sec.~\ref{sec:model}, we describe a general model for homodyne measurement of a multi-mode optomechanical system.
Measurement backaction arises from radiation pressure shot noise, which appears as a shared noise bath driving correlated diffusion of the oscillators.
A general set of quadrature estimators is defined in Sec.~\ref{sec:filters}, in terms of linear filters applied to the recorded homodyne signal.
Optimal quadrature filters, which minimize the estimator variance, are derived using linear regression, accounting for temporal correlations introduced by the oscillators' diffusive motion.

In Sec.~\ref{sec:statistics}, we derive the estimator covariance matrix, describing the imprecision added to the quadrature estimates by each source of noise in the recorded signal.
The estimator covariance can be measured using an ensemble of repeated measurements, which allows estimation of the covariance of a squeezed state, after correcting for the added noise covariance.

We derive an analytic approximation for the optimal single-oscillator filter in Sec.~\ref{sec:sql}, which minimizes the total added covariance.
The measurement strength, parameterized by the cooperativity $C_i$, quantifies the relative rate that information is gained from the system.
The optimal estimate, with sensitivity reaching the SQL, is obtained from a high-cooperativity measurement $\bathTemp_i + 1 \ll C_i \ll \omega_i / \Gamma_i$, which is bounded below by the thermal bath occupation $\bathTemp_i$ and above by the oscillator quality factor.
This condition ensures the measurement rate far exceeds the loss of information to the thermal environment.
The corresponding measurement backaction drives rapid diffusion of the oscillators, which is suppressed in the quadrature estimates by the appropriately optimized filters.

We demonstrate these matched-filter estimators on a simulated two-mode system in Sec.~\ref{sec:twomode}.
Simultaneous estimation of the state of multiple oscillators places additional constraints on the optimal estimate sensitivity, explored in Sec.~\ref{sec:twomode-sql}.
We show that the state of multiple oscillators can be retrodicted from the measurement record with precision at the SQL, if their frequencies are resolved by many linewidths.

Experimental results of matched filter estimates are presented in Sec.~\ref{sec:data}, obtained from a recent demonstration of the negative-mass instability between collective atomic spin and motion~\cite{Kohler2018}. Additional experimental complications from shot-to-shot fluctuations of system parameters had to be included in the model to recover accurate retrodicted estimates.
Finally, conclusions and outlook are summarized in Sec.~\ref{sec:conclusion}.

For clarity of notation throughout, vectors will be notated in bold (e.g. $\vec{v}$), and matrices in Roman typeface (e.g. $\mat{M}$).
Hermitian amplitude and phase quadratures of bosonic operators, such as $\cavHat$, are defined according to
\begin{align}
\label{eq:quadrature-def}
\cavAM & = \frac{1}{\sqrt{2}} \bigl( \cavDag + \cavHat \bigr) 
& &
\text{and}
&
\cavPM & = \frac{i}{\sqrt{2}} \bigl( \cavDag - \cavHat \bigr) \text{,}
\end{align}
respectively.
In particular, the quadratures of the optomechanical oscillators will be notated as a generalized position and momentum,
\begin{align}
\hat{X}_i & = \frac{1}{\sqrt{2}} \bigl( \adag_i + \ahat_i \bigr) 
& &
\text{and}
&
\hat{P}_i & = \frac{i}{\sqrt{2}} \bigl( \adag_i - \ahat_i \bigr) \text{,}
\end{align}
respectively.

\section{Simultaneous optomechanical measurement}
\label{sec:model}

Consider an ensemble of $\Nosc$ harmonic oscillators, illustrated in Fig.~\ref{fig:simulation}a, described by bosonic operators $\ahat_i$ evolving at frequencies $\omega_i$, which are dispersively coupled to a driven single-mode optical cavity with independent linear optomechanical coupling strengths $g_i$~\cite{Aspelmeyer2014}.
The cavity resonance frequency is shifted by the sum of the oscillators' displacements, which can be continuously measured by driving the cavity on resonance, such that the oscillators' motion modulates the phase quadrature of reflected light and is recorded using optical homodyne detection.

For small displacements, the dispersive shift of the cavity frequency is small relative to the cavity linewidth ${ \sum_i g_i \langle \adag_i + \ahat_i \rangle \ll \kappa }$, and the dynamics of the cavity field can be linearized in terms of fluctuations $\cavHat$ around an average cavity photon number $\nbar$, in a frame rotating at the cavity drive frequency $\probeW$, yielding a multi-mode generalization of the linearized optomechanical Hamiltonian~\cite{Aspelmeyer2014}
\begin{multline}
\label{eq:hamiltonian}
\mathcal{H} = - \hbar \Dpc \cavDag \cavHat + \sum_i \hbar \omega_i \adag_i \ahat_i
 + 2 \sum_i \hbar \sqrt{\nbar} g_i \cavAM \hat{X}_i
\end{multline}
where $\Dpc = \probeW - \cavityW$ is the detuning between the drive and cavity resonance frequency $\cavityW$, having dropped constant energy terms.

In order to simultaneously measure the intrinsic dynamics of multiple independent oscillators -- without introducing optically mediated coupling~\cite{Shkarin2014,Spethmann2015}, spring shifts~\cite{Sheard2004,Corbitt2006}, or damping~\cite{Arcizet2006,Gigan2006,Schliesser2006} -- we always consider a resonantly driven cavity, with $\Dpc = 0$.
The Heisenberg-Langevin equation of motion for the state of the cavity field $\cavHat(t)$
\begin{align}
\label{eq:cavity-eom}
\dot{ \cavHat } = -  i  \sum_i \sqrt{2 \nbar} g_i \hat{X}_i - \kappa \cavHat + \sqrt{2 \kappa} \vac
\end{align}
is obtained from Eq.~\eqref{eq:hamiltonian}, with the addition of input and output terms~\cite{Gardiner1985} introducing vacuum fluctuations $\vac$  from optical coupling to the environment, parametrized by the cavity half-linewidth $\kappa$.

For simultaneous measurement of multiple oscillators, it is advantageous to work in the fast-cavity (unresolved-sideband) regime defined by $\kappa \gg \omega_i$, such that the cavity field is sensitive across a wide bandwidth to dynamics of many oscillators at well-resolved frequencies.
Eq.~\eqref{eq:cavity-eom} can then be solved under the adiabatic approximation $\dot{ \cavHat } \approx 0$, assuming the cavity field equilibrates to the oscillators' motion nearly instantaneously, yielding solutions for the amplitude and phase quadratures
\begin{subequations}
\begin{align}
\label{eq:cavity-am}
\cavAM(t) & = \sqrt{\frac{2}{\kappa}} \vacAM(t)
\text{ and}
\\
\label{eq:cavity-pm}
\cavPM(t) & = \frac{2 \sqrt{\nbar} }{\kappa} \sum_i g_i \hat{X}_i(t) + \sqrt{\frac{2}{\kappa}} \vacPM(t)
\text{,}
\end{align}
\end{subequations}
respectively.
The cavity input fluctuations are assumed to be in the vacuum state, described by the two-time correlation $\langle \vac(t) \vacdag(t') \rangle = \delta(t-t')$.

The optical field leaking out of the cavity, determined by the boundary condition
$\sqrt{2 \kappa} \cavHat = \out - \vac$, carries information about the oscillator dynamics in its phase quadrature.  
The optical phase is recorded using a balanced homodyne detector, resulting in a signal proportional to the instantaneous shift of the cavity frequency due to the displacement of each oscillator
\begin{align}
\label{eq:signal-general}
\sig(t) = \sqrt{2} \sum_i g_i \hat{X}_i(t) + \sqrt{ \Psn } \vac_\text{SN}(t)
\text{.}
\end{align} 
The second term describes the added measurement shot noise, due to vacuum fluctuations of the optical probe, with two-time correlation $\langle \vac_\text{SN}(t) \vac_\text{SN}(t') \rangle = \delta(t - t')$ and normalized shot noise PSD $\Psn = \kappa/(8 \epsilon \nbar)$ in terms of the total cavity photon detection efficiency $\epsilon$.

\subsection{Derivation of oscillator trajectories}

Accurate retrodiction of the initial state of the oscillators from their subsequent evolution requires knowledge of the system's coherent dynamics, in addition to a complete stochastic model for all incoherent noise sources.
The oscillator equations of motion derived from Eq.~\eqref{eq:hamiltonian} are given by
\footnote{ assuming $g_i > 0$, without loss of generality. }
\begin{align}
\label{eq:osc-eom}
\dot{ \ahat }_i = (- \Gamma_i/2 - i \omega_i) \ahat_i - i \sqrt{ C_i \Gamma_i } \vacAM  + \sqrt{\Gamma_i} \bath_i 
\text{,}
\end{align} 
assuming each oscillator is intrinsically coupled with strength $\Gamma_i$ to an independent Markovian bath $\bath_i$, with two-time correlation $\langle \bath_i(t) \bathdag_i(t') \rangle = (\bathTemp_i + 1) \delta(t-t')$ parametrized by the equilibrium thermal occupation $\bathTemp_i$.

Radiation-pressure forces introduce measurement backaction, by coupling each oscillator to the cavity amplitude fluctuations $\cavAM$ described by Eq.~\eqref{eq:cavity-am}.
The optomechanical cooperativity 
\begin{align}
\label{eq:cooperativity}
C_i = 4 \nbar g_i^2 / \kappa \Gamma_i
\text{,}
\end{align}
which parametrizes the measurement strength, quantifies here the added equilibrium occupation due to diffusion from measurement backaction.
This backaction noise represents a common-mode bath, driving correlated diffusion of each oscillator during measurement~\cite{Spethmann2015}.

Each oscillator's trajectory is found by solving Eq.~\eqref{eq:osc-eom}, simulated numerically in Fig.~\ref{fig:simulation}c, and is readily separated into two parts:
\begin{align}
\label{eq:oscillator-position}
\hat{X}_i(t) = \hat{ \quadVec }^\Tr_i \vec{r}_i(t) + \hat{\sigDiff}_i(t)
\text{,}
\end{align} 
coherent evolution of the initial phase space quadratures, summarized by the two-element vector $ \hat{ \quadVec }_i = \begin{pmatrix} \hat{X}_i(0), \hat{P}_i(0) \end{pmatrix}^\Tr$, and accumulated incoherent diffusion $\hat{\sigDiff}_i(t)$.
The coherent state evolution is described by a vector of quadrature impulse response functions
\begin{align}
\label{eq:oscillator-response}
\vec{ r }_i(t) = e^{-\Gamma_i t/2} \begin{pmatrix} 
\cos \omega_i t \\
\sin \omega_i t
\end{pmatrix} \Theta(t)
\text{,}
\end{align} 
in terms of the Heaviside step function $\Theta(t)$.
The accumulated oscillator diffusion is given by the convolution
\begin{align}
\label{eq:oscillator-diffusion}
\hat{\sigDiff}_i(t) = \int_{0}^{\infty} d\tau \, \hat{ \vec{\noiseQuad} }^\Tr_i(\tau) \vec{r}_i(t - \tau)
\text{,}
\end{align}  
of the oscillator's response with the stochastic input noise, summarized by the input quadrature vector
\begin{align}
\label{eq:noise-drive}
\hat{ \vec{\noiseQuad} }_i(\tau) = 
\sqrt{\Gamma_i} 
\begin{pmatrix}
\bathAM_{i} (\tau)
\\
\bathPM_{i} (\tau)
- \sqrt{2 C_i} \vacAM (\tau)
\end{pmatrix}
\text{.}
\end{align} 

Substituting the oscillator trajectory given by Eq.~\eqref{eq:oscillator-position} into Eq.~\eqref{eq:signal-general}, the quantum mechanical model for the measured homodyne signal can be written as
\begin{align}
\label{eq:signal-solution}
\sig(t) = \sum_i \sqrt{2} g_i \! \left[ \hat{\quadVec }_i^\Tr \vec{ r }_i(t)  + \hat{\sigDiff}_i(t) \right]
 + \sqrt{\Psn} \, \vac_\text{SN}(t)
 \text{,}
\end{align} 
 which is a sum of the coherent ringdown of the initial states $\hat{ \quadVec }_i$, the accumulated diffusion of each oscillator during measurement, and measurement shot noise.
The incoherent part of Eq.~\eqref{eq:signal-solution} determines the total noise PSD of the homodyne signal, 
displayed for simulated signals in Fig.~\ref{fig:simulation}b, and contains thermal- and backaction-driven optomechanical responses in addition to the broadband shot noise floor.

Although these results are derived in the unresolved-sideband limit, Eq.~\eqref{eq:signal-solution} can be generalized for any resonantly driven cavity, provided only that $ \kappa \gg \Gamma_i $, by accounting for a reduced effective coupling strength $g_i \rightarrow g_i  \kappa / \sqrt{ \kappa^2 + \omega_i^2 }$, due to suppression of shot noise fluctuations and the optomechanical response by the cavity susceptibility, and an effective phase delay of the optically measured oscillator amplitude $\ahat_i \rightarrow e^{i \phi_i} \ahat_i$ with $\tan \phi_i = \omega_i/\kappa$.

\section{Linear filter estimation}
\label{sec:filters}

The goal of retrodiction, considered here, is to estimate the state of all oscillators at time $t=0$, represented by the quadrature vectors $ \hat{ \quadVec }_i $, from continuous measurement of their subsequent dynamics~\cite{Gammelmark2013}.
The homodyne photocurrent is amplified electronically and then digitally sampled, resulting in a classical recorded signal that contains noise arising from measurement shotnoise and quantum backaction.
For the linear systems considered in this work, such state estimation from the recorded signals can be approached as an essentially classical signal-filtering problem \cite{Muller-Ebhardt2009}.

A general set of linear filters applied to the observed homodyne signal is defined as
\begin{align}
\label{eq:general-filter}
\rawVec = \int_{0}^{\tf} dt \, \vec{m}(t) \sig(t)
\text{,}
\end{align} 
in terms of a vector of real-valued temporal weight functions $\vec{m}(t)$.
Here, the vector of filter outputs $\rawVec$ represent projections from the infinite-dimensional space of the continuous signal $\sig(t)$ onto temporal modes defined by the filter functions $\vec{m}(t)$.

Assuming all noise sources in Eq.~\eqref{eq:signal-solution} have zero mean, the average filter outputs can be directly evaluated and expressed as a matrix equation
\begin{align}
\label{eq:general-filter-mean}
\langle \rawVec \rangle & = \mat{J} \langle \hat{ \quadVec } \rangle
\end{align}
in terms of the $2 \Nosc$-element vector of initial quadrature amplitudes $\hat{ \quadVec }$ and $2 \Nosc \times 2 \Nosc $ normalization matrix $\mat{J}$
\begin{align}
\hat{ \quadVec } & = \begin{pmatrix}
\hat{ \quadVec }_1 \\ \hat{ \quadVec }_2 \\ \vdots \\ \hat { \quadVec }_\Nosc \end{pmatrix} 
&
\mat{J} =
\left(
\begin{array}{cccc}
\mat{J}_{11} & \mat{J}_{21} & \dots & \mat{J}_{\Nosc 1} 
\\
\mat{J}_{12} & \mat{J}_{22} & \dots & \mat{J}_{\Nosc 2}
\\
\vdots & \vdots & \ddots & \vdots
\\
\mat{J}_{1\Nosc} & \mat{J}_{2\Nosc} & \dots & \mat{J}_{\Nosc \Nosc}
\end{array}
\right)
\end{align}
defined by concatenation of the individual 2-dimensional quadrature vectors and $2 \times 2$ block matrices
\begin{align}
\label{eq:normalization-matrix}
\mat{J}_{ij} = \sqrt{2} g_j \int_{0}^{\tf} dt \, \vec{m}_i(t) \vec{ r }^\Tr_j(t)
\end{align} 
for 2-element vectors of filter functions $\vec{m}_i(t)$ defined for each oscillator $i$.

As an initial example, the filters can be defined according to each oscillator's quadrature response functions
\begin{align}
\label{eq:ols-filters}
\vec{m}^\text{OLS}_i(t) \equiv \vec{r}_i(t)
\text{.}
\end{align} 
For this choice of filters functions, the estimator results represent the projection of the signal onto each quadrature's coherent response function.
The diagonal elements of the matrix $\mat{J}$ describe the normalization of each filter and the off-diagonal elements reflect the non-orthogonality between quadrature responses, due to spectral overlap from the finite oscillator linewidths. 

This overlap between quadrature filters introduces spurious correlations in the raw filter outputs $\est{q}_i$.
Provided there are $2 \Nosc$ linearly independent filter functions $\vec{m}(t)$ that span the space of the quadrature response functions $\vec{r}_i(t)$, then Eq.~\eqref{eq:general-filter-mean} can be solved to recover the average initial quadrature amplitudes by inverting the normalization matrix $\mat{J}$.

By extension, a complete set of unbiased quadrature estimators can, therefore, be defined as
\begin{align}
\label{eq:estimator-definition}
\est{\quadVec} = \begin{pmatrix}
\est{\quadVec}_1 \\ \est{\quadVec}_2 \\ \vdots \\ \est{\quadVec}_\Nosc
\end{pmatrix} \equiv \mat{J}^{-1} \int_{0}^{\tf} dt \, \vec{m}(t) \sig(t)
\text{,}
\end{align} 
satisfying $  \langle \est{\quadVec}_i \rangle = \langle \hat{\quadVec}_i \rangle $.

\begin{figure}[tb!]
\includegraphics[width=3.375in]{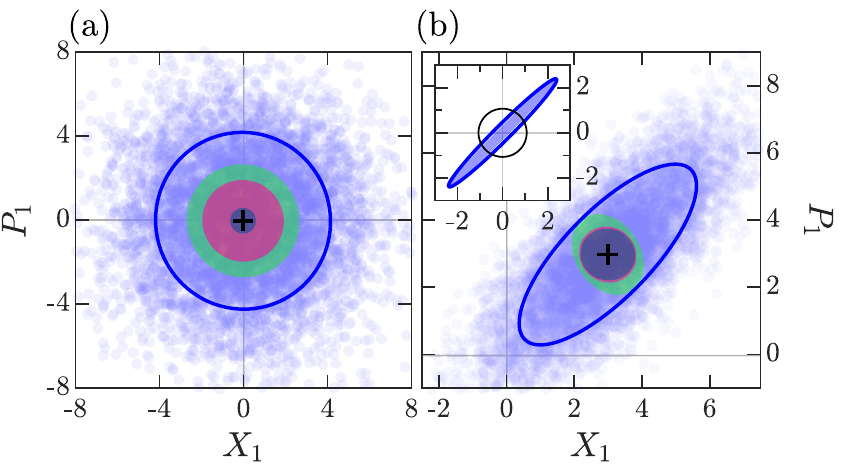}
\caption{
	\label{fig:single-oscillator}
	(a) Initial state estimates (blue points) from applying OLS filters to 8000 simulated signals for measurement of an oscillator in equilibrium with its thermal bath ($\omega_1 = 2\pi \times 125$ kHz, $\Gamma_1 = 2\pi \times 2$ kHz, and $\bathTemp_1 = 1$), with measurement cooperativity $C_1 = 3$. 
	The covariance of quadrature estimates defines a $68\%$ confidence ellipse (blue), which is the cumulative sum of the added measurement shot noise (black circle), thermal bath noise (red annulus), quantum backaction (green annulus), and the retrodicted state covariance.
	(b) Phase space distribution for simulated measurements of a displaced $-10$~dB squeezed vacuum state of the oscillator in (a), with measurement cooperativity $C_1 = 20$ and estimated using optimal filters derived from Eq.~\eqref{eq:gls-filters}. 
	(inset) Squeezing of the oscillator's initial state can be inferred after subtracting the added noise covariances (blue ellipse), revealing an initial quadrature variance below the zero-point scale (black circle).
}
\end{figure}

Results obtained by applying these filter estimators to simulated measurements of a single oscillator's trajectory are shown in Fig.~\ref{fig:single-oscillator}a.
The notation $\est{ \quadVec }$ is used here to indicate an estimator for the vector of quadrature operators $\hat{ \quadVec }$, corresponding to a temporal mode of the detected optical field.
The distribution of measured samples, obtained by application of Eq.~\eqref{eq:estimator-definition} to the recorded homodyne traces, can be described by the statistics of the thermal and quantum noise contained in the estimator model defined by Eqs.~\eqref{eq:signal-solution} and \eqref{eq:estimator-definition}.

\subsection{Generalized least-squares optimization}

The optimal set of filter functions $\vec{m}_i(t)$, which provide a minimum-variance unbiased estimate for any linear combination of quadratures $\hat{ \quadVec }$, can be derived using the method of Least Squares, such as commonly used for linear curve-fitting.
The filters defined by Eq.~\eqref{eq:ols-filters} are obtained from an Ordinary Least Squares (OLS) linear regression, by minimizing the sum of square residuals between the measured signal and the coherent model
\begin{align}
\Phi^\text{OLS}[\vec{m}(t)] \equiv \int_0^{\tf} dt\, \Bigl[ 
	\sig(t) - \sqrt{2} \sum_j g_j \est{ \quadVec }^\Tr_j \vec{ r }_j(t) 
\Bigr]^2
\text{,}
\end{align}
parameterized by estimators $\est{\quadVec}_j$.

However, the Gauss-Markov theorem~\cite{McElroy1967} proves that these estimators are optimal only when the signal noise is temporally uncorrelated (white noise).
Diffusive motion of the oscillators during the measurement, driven by thermal and backaction fluctuations, generates temporal correlations in the signal noise, indicated by structure in the PSD shown in Fig.~\ref{fig:simulation}b.
In the presence of temporally correlated noise, a minimum variance unbiased estimator can be constructed through linear regression using the Generalized Least Squares (GLS) method~\cite{Aitken1936}.
This method can be understood conceptually as de-correlating the temporal signal by inverting the known two-time noise correlation function prior to performing linear regression.

The inverted noise correlation function is more easily defined for a signal sampled at discrete times $t_n = n  / f_s$, with sample frequency $f_s$ and count $\Ntime = f_s \tf$, which is typical for most experimental applications.
The discrete two-time correlation function of the added diffusive and measurement noise defines the $\Ntime \times \Ntime$ square matrix
\begin{align}
\label{eq:twotime-noise-matrix}
\mat{ \Omega }_{nm} = 2 \sum_{kl} g_k g_l \langle \hat{\sigDiff}_k(t_n) \hat{\sigDiff}_l(t_m) \rangle + \Psn f_s \, \delta_{nm}
\text{,}
\end{align} 
which can be inverted numerically to derive the discrete GLS filter functions
\begin{align}
\label{eq:gls-filters}
\vec{m}^\text{GLS}(t_n) \equiv \sum_{m=0}^{\Ntime-1} [\mat{ \Omega }^{-1}]_{nm} \vec{r}(t_m)
\text{.}
\end{align} 
These filters implicitly transform the signal to de-correlate the noise, recovering conditions to satisfy the Gauss-Markov theorem and, therefore, providing the minimum-variance unbiased estimate for any general optomechanical measurement.

The effect of diffusive noise on the optimal filters is further illustrated in Sec.~\ref{sec:sql}, where an analytic formula is derived that approximates the optimal filters for a single oscillator.
The numerical method defined in Eqs.~\eqref{eq:twotime-noise-matrix} and \eqref{eq:gls-filters}, however, facilitates calculation of optimal filters in any conditions and is fully general for state estimation with multiple oscillators, considered in Sec.~\ref{sec:twomode}.

\section{Linear estimator statistics}
\label{sec:statistics}

In addition to providing an estimate of the mean quadrature amplitudes $\langle \hat{ \quadVec }_i \rangle$, an ensemble of estimates from repeated measurements of identically prepared states can be used to estimate the multi-mode quadrature covariance of the initial state
\begin{align}
\operatorname{cov}[\hat{ \quadVec }] \equiv \Re[ \langle \hat{ \quadVec } \hat{ \quadVec }^\Tr \rangle ] - \langle \hat{ \quadVec } \rangle \langle \hat{ \quadVec }^\Tr \rangle
\text{.}
\end{align} 

Noise in the measured signal introduces errors into the quadrature estimates, increasing the uncertainty of the retrodicted state.
For linear filters, the additive noise in the signal described by Eq.~\eqref{eq:signal-solution} introduces a systematic bias to the covariance of the observed quadrature estimates
\begin{align}	
\label{eq:estimate-covariance}
\mat{ \Sigma} \equiv \operatorname{cov}[\est{ \quadVec }]
=  \operatorname{cov}[\hat{ \quadVec }] + \matTh + \matBA + \matSN
\text{,}
\end{align} 
which is a sum of the actual state covariance $\operatorname{cov}[\hat{ \quadVec }]$ and the additive covariance from each independent noise source -- thermal diffusion $\matTh$, quantum backaction $\matBA$, and measurement shot noise $\matSN$ -- indicated by shaded regions in Fig.~\ref{fig:single-oscillator}.

The estimate covariance added by measurement shot noise is given by
\begin{align}
\label{eq:shotnoise-covariance}
\matSN & = \Psn \,
\int_{0}^{\tf}dt \,  \mat{J}^{-1} \vec{m}(t)  
\left(  \mat{J}^{-1} \vec{m} (t) \right)^{\Tr}
\text{,}
\end{align} 
in terms of the overlap matrix between each pair of normalized quadrature filters.
The filter functions of interest are generally not orthogonal, therefore measurement shot noise will induce correlated errors in the quadrature estimates, due to the spectral overlap between pairs of filters, described by non-zero off-diagonal components of $\matSN$.

Diffusion of the oscillators' states during the measurement, driven by their intrinsic thermal baths as well as quantum backaction, also adds to both the variances and covariances of the quadrature estimates.
The response of oscillator $k$ to a generic bath fluctuation at time $\tau$ has an integrated effect on the quadrature estimates, described by the $2 \Nosc \times 2$ matrix-valued function
\begin{align}
\label{eq:generic-diffusive-noise}
\mat{N}_k(\tau) = \sqrt{2 \Gamma_k} g_k \mat{J}^{-1} \int_0^\tf dt \, \vec{m}(t) \vec{r}^\Tr_{k}(t - \tau)
\text{,}
\end{align} 
which arises from applying the estimator defined by Eq.~\eqref{eq:estimator-definition} to the diffusion term of the full signal model in Eq.~\eqref{eq:signal-solution}.
The thermal baths of each oscillator are assumed to be independent, so that the total thermal noise covariance is simply given by a sum over the variance of induced estimate perturbations, weighted by each oscillator's bath occupation
\begin{align}
\label{eq:thermal-covariance}
\matTh & = \sum_{k} \Bigl( \bathTemp_k + \frac{1}{2} \Bigr) \,
\int_0^\tf d\tau \, \mat{N}_k(\tau) \mat{N}^\Tr_k(\tau)
\text{.}
\end{align} 

The quantum backaction, however, induces correlated diffusion of the oscillators during the measurement, since the oscillators all respond to the same amplitude fluctuations of the cavity field and have finite spectral overlap of their susceptibilities (assuming non-zero oscillator linewidths).
Diffusive motion from this common optical bath induces correlated errors in the quadrature estimates, with covariance given by a sum over all oscillator pairs
\begin{align}
\label{eq:backaction-covariance}
\matBA & =
\sum_{kl} \sqrt{C_k C_l } \,
\int_0^\tf d\tau \, \mat{N}_k(\tau)
\begin{pmatrix}
0 & 0 \\ 0 & 1
\end{pmatrix}
 \mat{N}^\Tr_l(\tau)
\\
& \approx
\sum_{kl} \frac{ \sqrt{C_k C_l } }{2} \,
\int_0^\tf d\tau \, \mat{N}_k(\tau) \mat{N}^\Tr_l(\tau)
\text{,}
\end{align} 
assuming in the last line that $\omega_k + \omega_l \gg \Gamma_k + \Gamma_l$.

As defined above, the added noise covariance matrices $\matTh$, $\matBA$, and $\matSN$ are expressed in units of an equivalent thermal phonon occupation and represent the measurement uncertainty for any single estimate obtained from these filters.
For a given measurement configuration, the optimal filters minimize this added noise covariance and provide estimates of the quadrature amplitudes with the least uncertainty.

The quadrature covariance of the initial multi-mode state can be inferred from an ensemble of repeated measurements, by inverting Eq.~\eqref{eq:estimate-covariance}
\begin{align}
\operatorname{cov}[\hat{ \quadVec }] = \mat{\Sigma} - \matTh - \matBA - \matSN
\text{,}
\end{align}
assuming identical preparation of the initial state for each measurement.
If the system and bath parameters are independently calibrated, then the bias matrices $ \matTh $, $ \matBA $, and $ \matSN $ can be precisely calculated and subtracted to recover the inferred multi-mode state covariance.
The statistical uncertainty of the inferred covariance is then limited by the uncertainty of the estimator covariance $\mat{\Sigma}$, which can be reduced by minimizing the total added noise covariance $\matTh + \matBA + \matSN$ for each sample, in addition to increasing the sample size $\Nsamp$ (see Appendix~\ref{app:covariance-statistics}).

Fig.~\ref{fig:single-oscillator}b demonstrates retrodiction of an oscillator initially prepared in a squeezed state $\ket{ \zeta } = \hat{\mathcal{S}}(\zeta) \ket{ 0 }$, defined by the single-mode squeezing operator~\cite{WallsAndMilburn}
\begin{align}
\hat{ \mathcal{S} }(\zeta) = \exp \frac{1}{2} \bigl( \zeta^* \ahat_1^2 - \zeta  \ahat_1^{\dagger 2} \bigr)
\text{.}
\end{align}
Squeezing of the quadrature variance below the ground-state zero-point-motion cannot be directly observed with the phase-independent quadrature estimates considered in this work, since they are constrained by the SQL.
However, quadrature squeezing can be inferred from the covariance of an ensemble of estimates, demonstrated in Fig.~\ref{fig:single-oscillator}b, only after subtracting the added noise covariance matrices, which are numerically evaluated from Eqs.~\eqref{eq:shotnoise-covariance}, \eqref{eq:thermal-covariance}, and \eqref{eq:backaction-covariance} as described in Appendix~\ref{app:response-correlations}.

\section{Standard quantum limit for retrodiction}
\label{sec:sql}

\begin{figure}[tb!]
\includegraphics[width=3.375in]{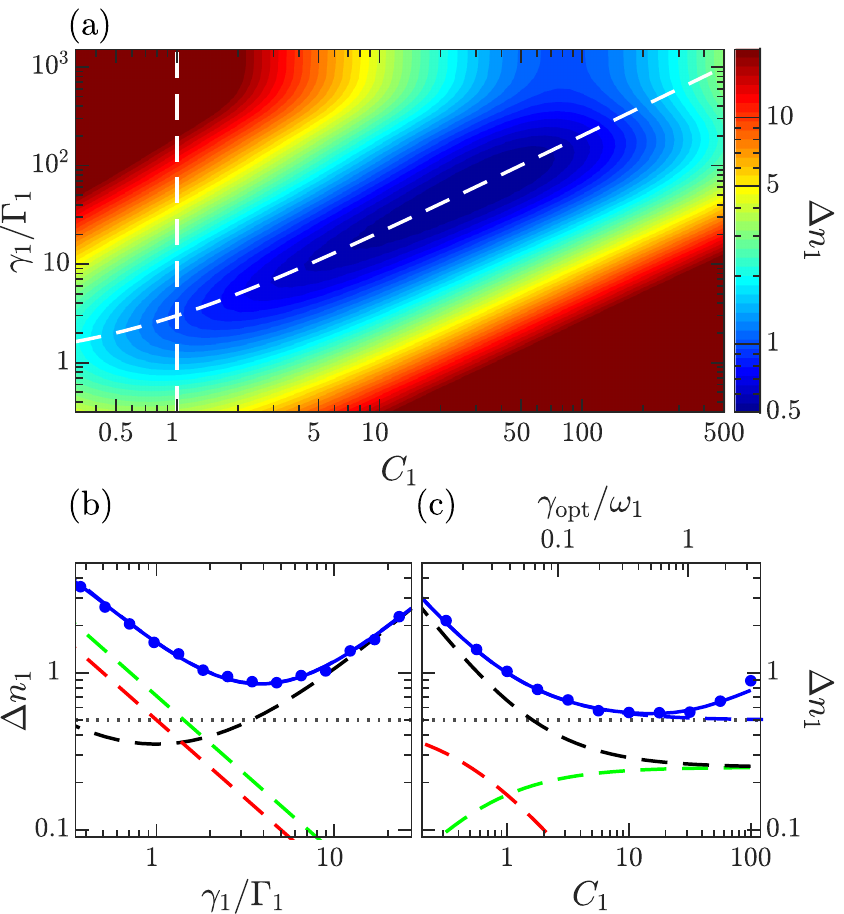}
\caption{
	\label{fig:filter-sql}
	(a) Total estimate imprecision for retrodiction of a single oscillator's state as a function of the measurement cooperativity $C_1$ and exponential filter decay rate $\gFilt_1$, calculated for the same oscillator as Fig.~\ref{fig:single-oscillator} but assuming a zero-temperature bath $\bathTemp_1 = 0$ and ideal detection efficiency $\epsilon = 1$.
	Dashed white lines mark the line cuts plotted in (b) and (c).
	(b) Total added noise for various exponential filter decay rates $\gFilt_1$ at a fixed cooperativity $C_1=1$, from numerical calculations (solid blue line) and 4000 simulated estimates (blue dots).	
	Analytic approximations for each noise component (dashed black: shot noise, green: backaction, red: thermal) illustrate that thermal- and backaction-driven diffusion shift the optimal filter to a faster decay rate.
	(c) The minimized added noise for the optimal exponential filter approaches the SQL (dotted gray line) in the backaction-dominated regime $C_1 \gg 1$.
	When $\gFilt_\text{opt} \gtrsim \omega_1$, the filters no longer provide equal information about both quadratures and the analytic approximation (dashed lines) deviates from the full numerical calculation (solid blue line).
	Simulated estimates with the GLS filters derived from  Eq.~\eqref{eq:gls-filters} (blue dots) reach identical noise limits, provided $\gFilt_1 \gg \tf^{-1}$.
}
\end{figure}

Realizing optimal state retrodiction of a given ensemble of oscillators involves two separate choices. 
First, the intra-cavity intensity $\nbar$ must be chosen at the time of measurement, determining the measurement cooperativity $C_i$ for each oscillator.
Second, for a given measurement strength, the optimal filter must be derived to obtain the best state estimates from each recorded trace, provided by Eq.~\eqref{eq:gls-filters}.
In this section, we consider retrodiction of the state of a single oscillator and determine the measurement conditions for achieving sensitivity at the SQL.

The OLS filters defined in Eq.~\eqref{eq:ols-filters} weight the filter estimate according to the coherent decay of the initial state.
However, diffusion of the oscillator state after $t=0$ adds noise to the observed trajectories, illustrated in Fig.~\ref{fig:simulation}c, accumulating at a rate proportional to the measurement cooperativity $C_1$ and the thermal bath occupation $\bathTemp_1$.
An optimal filter should appropriately weight the relative signal to noise at each subsequent time $t$, implying that the optimal filter envelope must decay faster than the coherent response functions $\vec{r}_1(t)$.

The GLS filters derived from Eq.~\eqref{eq:gls-filters} can be used to calculate the optimal filter for any particular system parameters, however an analytic model yields further intuition about the optimal measurement cooperativity.
For estimation of a single oscillator, the GLS filters are well-approximated by a set of exponentially damped sinusoidal filters
\begin{align}
\label{eq:exp-filters}
\vec{m}^\text{exp}_1(t) \equiv e^{-\gFilt_1 t/2}
\begin{pmatrix}
\cos \omega_1 t \\ \sin \omega_1 t
\end{pmatrix}
\text{,}
\end{align}
parametrized by an arbitrary exponential decay rate $\gFilt_1$, for sufficiently long observation intervals $\Gamma_1, \gamma_1 \gg \tf^{-1}$.
In general, the total noise added to the quadrature estimates for any oscillator $i$ can be quantified by the average variance
\begin{align}
\label{eq:osc-imprecision}
\imp_i = \frac{1}{2} \operatorname{Tr}\left[ \matTh_{ii} + \matBA_{ii} + \matSN_{ii} \right]
\text{,}
\end{align}
where $\matTh_{ii}$ refers to the $2 \times 2$ diagonal block matrix corresponding to oscillator $i$.
For estimation of a single oscillator with the exponential filters, this added noise occupation can be numerically evaluated for a given measurement cooperativity $C_1$ and filter decay rate $\gFilt_1$, with results summarized in Fig.~\ref{fig:filter-sql}a.

The individual noise covariance matrices given by Eqs.~\eqref{eq:shotnoise-covariance}, \eqref{eq:thermal-covariance}, and \eqref{eq:backaction-covariance} can also be evaluated analytically and shown to be approximately proportional to the identity matrix, for filters that decay slower than the oscillation frequency $\gamma_1 \ll \omega_1$
 and assuming a measurement duration that is sufficiently long to capture the full transient response $\Gamma_1, \gamma_1 \gg \tf^{-1}$.
The uncertainty contributed from each noise source, therefore, can be fully described by the added quadrature variance, expressed as equivalent added thermal occupation
\begin{subequations}
\begin{align}
\imp_1( \matTh ) & \equiv \frac{1}{2} \operatorname{Tr}\left[ \matTh_{11} \right] 
\approx \Bigl( \bathTemp_1 + \frac{1}{2} \Bigr) \frac{\Gamma_1}{\gFilt_1}
\text{,}
\\
\imp_1(\matBA) & \equiv \frac{1}{2} \operatorname{Tr}\left[ \matBA_{11} \right] 
\approx \frac{C_1}{2}  \frac{\Gamma_1}{\gFilt_1}
\text{,}
\\
\imp_1(\matSN)  & \equiv \frac{1}{2} \operatorname{Tr}\left[ \matSN_{11} \right] 
\approx \frac{1}{2 \epsilon C_1} \frac{(\Gamma_1 + \gFilt_1)^2}{4 \Gamma_1 \gFilt_1}
\text{.}
\end{align}
\end{subequations}
These three expressions sum to give the total added noise, as displayed in Fig.~\ref{fig:filter-sql}b-c.

For measurements performed at a particular cooperativity $C_1$, the uncertainty of estimates obtained from the recorded traces can be evaluated individually for each noise source as a function of the filter decay rate $\gFilt_1$, as shown in Fig.~\ref{fig:filter-sql}b.
As predicted by the Gauss-Markov theorem~\cite{Rao1976}, the OLS estimators defined by Eq.~\eqref{eq:ols-filters}, equivalently $\gFilt_1 = \Gamma_1$, only minimize the added variance due to the temporally uncorrelated measurement noise $\imp_1(\matSN)$.
The OLS optimization does not minimize the estimator variance for temporally correlated noise~\cite{Rao1976}, driven by the thermal bath and quantum backaction, and the optimal filter decay rate is increased $\gFilt_1 > \Gamma_1$, because a shorter temporal filter envelope captures less of the accumulated diffusion at later times.

Minimizing the total added variance, the optimal exponential filter decay rate for a given measurement cooperativity $C_1$ is 
\begin{align}
\label{eq:optimal-filter-decay}
\gOpt = \Gamma_1 \sqrt{ 1 + 4 \epsilon C_1 (C_1 + 2 \bathTemp_1 + 1) }
\text{.}
\end{align}
The added noise for this optimized exponential filter is plotted in Fig.~\ref{fig:filter-sql}c as a function of measurement cooperativity $C_1$.
The optimal measurement condition for retrodiction is achieved in the limit of high cooperativity, corresponding to backaction-dominated diffusion of the oscillator during the measurement.

In this high-cooperativity regime, where diffusion from backaction exceeds that from the oscillator's thermal motion (satisfying $C_1 \gg \bathTemp_1 + 1/2$), and where the effect of backaction in the measured signal exceeds that of measurement shot noise (satisfying $C_1 \gg \epsilon^{-1/2}$), the optimal filter decay rate is approximately
\begin{align}
\label{eq:diffusive-filter-decay}
\gFilt_\text{opt} \approx 2 \sqrt{\epsilon} C_1 \Gamma_1 = \frac{8 \sqrt{\epsilon} \nbar g_1^2}{\kappa}
\text{,}
\end{align}
with corresponding minimized total added noise occupation
\begin{align}
\label{eq:sql-noise}
n_\text{min} = \frac{1}{2 \sqrt{\epsilon}}
\text{.}
\end{align}
The SQL for state retrodiction, therefore, is reached in this backaction-dominated regime, assuming ideal detection efficiency $\epsilon = 1$.
The result in Eq.~\eqref{eq:diffusive-filter-decay} also defines the optimal filter envelope for an oscillator with negligible intrinsic damping $\Gamma_1 \rightarrow 0$.
In this limit, the optimal measurement sensitivity is realized at any finite measurement strength, provided an observation period much longer than the backaction diffusion timescale $\tf \gg 2\pi \kappa/(4 \nbar g_1^2)$.

In the strong-measurement limit $C_1 \rightarrow \infty$, this best exponential filter evolves toward a delta function, which would describe a fully projective, instantaneous measurement of the oscillator's position.
However, as observed in Fig.~\ref{fig:filter-sql}a,c, the optimal measurement cooperativity is bounded from above by the oscillator's quality factor $2 \sqrt{\epsilon} C_1 \ll \omega_1 / \Gamma_1$.
Beyond this bound, the filter envelope decays within an oscillation period, violating the approximation $\gFilt_1 \ll \omega_1$ above, and retrodiction no longer provides a phase-independent estimate of both oscillator quadratures.

It is noteworthy to contrast these imprecision limits for state retrodiction with those for the well-demonstrated limits for continuous displacement~\cite{Teufel2009,Anetsberger2010,Rossi2018} and force~\cite{Schreppler2014,Mason2018} detection.
In each case, the optimal measurement is obtained with an equal imprecision added by measurement noise and quantum backaction.
When an optomechanical oscillator is employed as a sensor for external forces or displacements, for instance from gravitational waves~\cite{LIGODetection}, then diffusion driven by quantum backaction increases the measurement imprecision at later times, and the optimal sensitivity on mechanical resonance is achieved with cooperativity $C_1 = (2 \sqrt{\epsilon})^{-1}$~\cite{Schreppler2014}, typically of unity order. 

However, for the case of retrodiction, the results described here indicate that the optimal sensitivity is reached in the high-cooperativity regime $C_1 \gg \bathTemp_1 + 1/2$, where information about the oscillator's initial state is rapidly extracted.
Measurement of the oscillator's state by the cavity mode inherently results in backaction noise added to the oscillator.
Nevertheless, it is preferable to increase the measurement cooperativity such that the measurement rate and the backaction-induced diffusion far exceed the loss of state information to the unmeasured modes of the oscillator's thermal bath.
The additional diffusive noise added to the oscillator's trajectory during measurement is suppressed in the estimate, by using an appropriately short filter profile.

\section{Two-mode state estimation}
\label{sec:twomode}

\begin{figure}[tb!]
\includegraphics[width=3.375in]{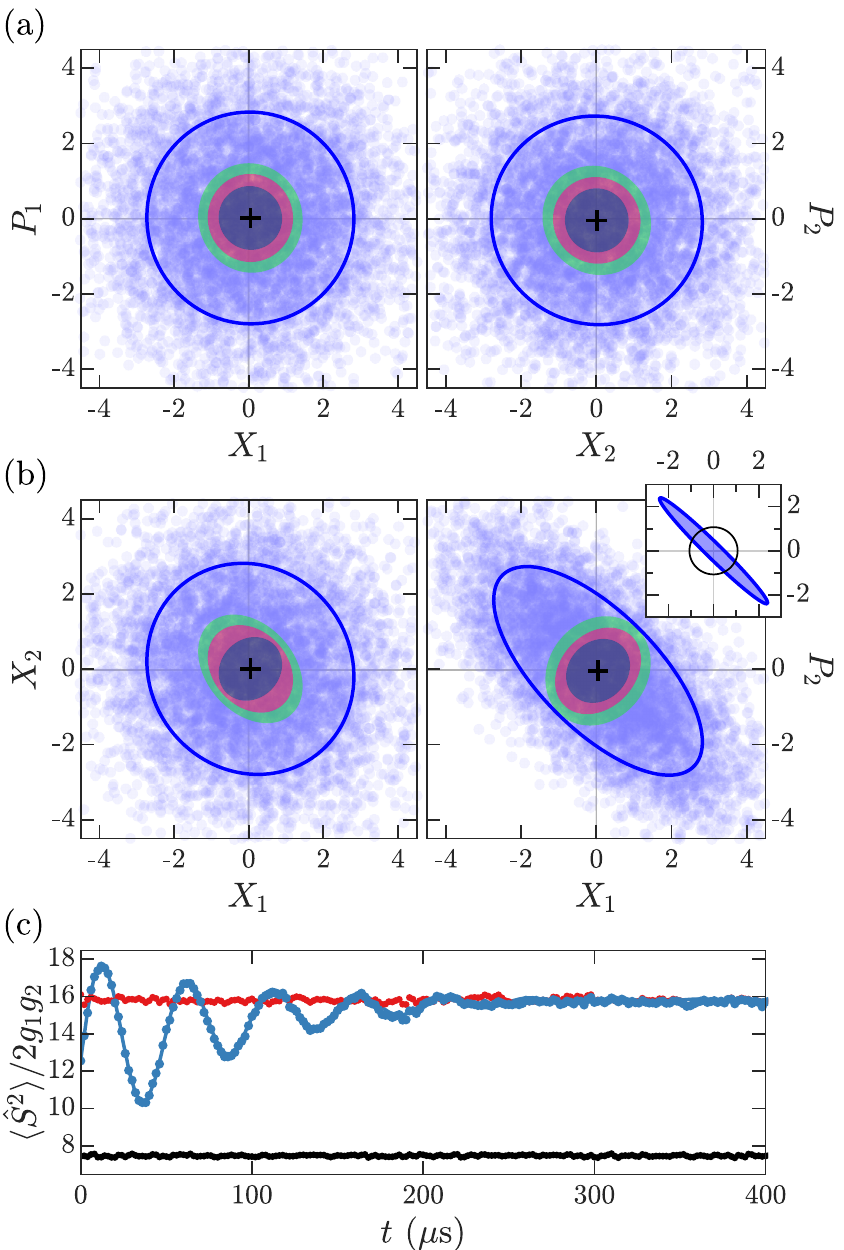}
\caption{
	\label{fig:two-mode}
	Retrodiction from 8000 simulated measurements of a $-10$~dB two-mode squeezed state $\ket{ z = 1.15 i }$ of the oscillators simulated in Fig.~\ref{fig:simulation}b, with measurement cooperativity $C_i = 5.3$ and bath occupation $\bathTemp_1 = \bathTemp_2 = 1$.
	(a) Independent phase space distributions for each oscillator show an effective thermal distribution of estimates, convolved with additive variance from measurement (black circle), thermal (red annulus), and backaction noise (green annulus).
	(b)
	For an imaginary squeezing parameter, the positions of the two oscillators are uncorrelated (left), however estimate errors are weakly correlated due to spectral overlap of the oscillator susceptibilities and filters.
	Correlations between out-of-phase quadratures of each oscillator (right) reveal the two-mode squeezed state, when corrected for the added covariance and correlation from all noise sources (inset).
	(c) 
	The mean square signal (blue dots) provides an independent signature of the two-mode correlations, unbiased by the added noise, which corroborates the inferred signal reconstructed by time evolution of the estimated two-mode covariance (blue line).
	The initial state transient decays to equilibrium with thermal and backaction noise (red dots) in addition to measurement shot noise (black dots).
}
\end{figure}

The formalism developed in Sec.~\ref{sec:model} and \ref{sec:filters} is already fully general for estimation of multi-mode states of $\Nosc$ oscillators.
Similar multi-mode estimation has been applied in experiments to obtain estimates for two-mode states \cite{Palomaki2013a,Spethmann2015,Kohler2018} and to demonstrate entanglement \cite{Palomaki2013} from correlations observed in subsequent measurements of two modes.

As an example, consider simultaneous retrodiction of two oscillators prepared in a two-mode squeezed state (TMSS)
$\ket{ z } = \hat{ \mathcal{S} }_2 (z) \ket{0,0}$,
generated from ground-state oscillators through the action of the two-mode squeezing operator~\cite{WallsAndMilburn}
\begin{align}
\hat{ \mathcal{S} }_2(z) = \exp \bigl( z^* \ahat_1 \ahat_2 -  z \adag_1 \adag_2 \bigr)
\text{.}
\end{align}
After the squeezing interaction is turned off, estimates of each quadrature of the two-mode system are obtained by applying optimized filters, calculated from Eqs.~\eqref{eq:twotime-noise-matrix} and \eqref{eq:gls-filters}, to the subsequently observed free transient decay.

The resulting $2 \Nosc$-dimensional Gaussian phase space distribution defines an ellipsoid in phase space, fully characterized by its mean and covariance.
The distribution can be visualized in terms of orthogonal 2-d projections, as shown in Fig.~\ref{fig:two-mode} for estimates obtained from simulated homodyne measurements of a TMSS.
Projections onto the 2-d phase space for each individual oscillator, shown in Fig.~\ref{fig:two-mode}a, are equivalent to tracing over the other oscillator's state and reflect an effective thermal occupation $\langle \adag_i \ahat_i \rangle = \sinh^2 |z|$. 

The presence of two-mode squeezing is revealed by correlations between quadratures of the different oscillators, which are displayed in Fig.~\ref{fig:two-mode}b.
`In-phase' correlations, between the positions of the two oscillators $\hat{X}_1(0)$ and $\hat{X}_2(0)$, are produced by the real part of $z$, while the imaginary part generates `out-of-phase' correlations, between the position of one oscillator and the momentum of the other, as shown in Fig.~\ref{fig:two-mode}b.
The other two orthogonal projections of the 4-d phase space are qualitatively similar, but not shown.
Once again, recovering the actual state covariance requires subtracting the covariances added by each noise source, which are numerically evaluated from independent knowledge of the system parameters.

\subsection{Verification of estimated correlation}

The two-mode covariance inferred from the matched-filter estimates can be experimentally validated through comparison to model-independent statistics of the recorded signals.
For a multi-mode system, the mean-squared homodyne signal also reveals information about the initial correlations~\cite{Kohler2018}
\begin{multline}
\label{eq:dsq}
\langle \sig^2(t) \rangle \approx 
	2 \sum_{ij} g_i g_j	\vec{r}_i^\Tr(t) \langle \hat{ \quadVec }_i \hat{ \quadVec }_j^\Tr \rangle \vec{r}_j(t)
	+ f_\text{BW} \Psn
	\\
	+ \sum_{ij} g_i g_j \Bigl[ ( 2 \bathTemp_i + 1 ) \delta_{ij} + \sqrt{C_i C_j} \Bigr] R_{ij}(t,t)
\text{,}
\end{multline}
again assuming $\omega_i + \omega_j \gg \Gamma_i + \Gamma_j$, where $f_{BW}$ is the bandwidth of the recorded signal.
This expression reflects coherent time evolution of the mean and covariance of the initial quadrature amplitudes, in addition to relaxation given by
\begin{align}
\label{eq:response-correlations}
R_{ij}(t,t') = \int_0^\infty d\tau \, \vec{r}^\Tr_{i}(t - \tau) \vec{r}_{j}(t' - \tau) 
\end{align}
to an equilibrium signal variance determined by the thermal baths and measurement backaction.

Correlations between oscillators appear in this signal as transient beatnotes at the sum and difference frequencies, unperturbed by the thermal, backaction, and measurement noise that biases the filter covariances.
This signal, therefore, serves as an independent statistic for comparison to the matched-filter estimates, through reconstructing the predicted beatnote for the inferred two-mode state, shown in Fig.~\ref{fig:two-mode}c.
This comparison indicates how faithfully the filter model matches the system dynamics and serves as a check for calibration of the added noise covariance that must be subtracted from the filter estimates.

The `out-of-phase' correlations for the TMSS simulated in Fig.~\ref{fig:two-mode} are characteristic of those generated by the negative-mass instability observed in Ref.~\cite{Kohler2018}, produced by resonant coupling between positive- and negative-mass oscillators.
For an effective negative-mass oscillator, the coherent state evolution corresponds to an opposite rotation in phase space, corresponding to $\omega_i \rightarrow - \omega_i$ in Eq.~\eqref{eq:osc-eom} and Eq.~\eqref{eq:oscillator-response}.
By consequence, for the TMSS represented in Fig.~\ref{fig:two-mode}c, the 2nd-order coherence $\langle \ahat_1 \ahat_2 \rangle$ evolves at the frequency difference, with amplitude and phase directly reflecting the magnitude and phase of correlations between the two oscillators.

\section{Limits of multi-mode estimate precision}
\label{sec:twomode-sql}

For a given multi-mode measurement record, with independently calibrated system and noise parameters, the GLS method facilitates numerical calculation of the optimal filter to recover the multi-mode state estimates.
For a single oscillator, the optimal estimate imprecision approaches the SQL in the limit of high cooperativity, as demonstrated in Sec.~\ref{sec:sql}.
However, diffusive motion of each oscillator in the measurement record introduces additional imprecision to the estimate results.
Here we explore the optimal two-mode measurement strength and additional limits to the estimate precision due to the presence of a second oscillator

Consider state retrodiction for simultaneous observation of two oscillators that differ only in their resonance frequency, with frequency separation $\delta = \omega_2 - \omega_1$.
The distinguishability of the two oscillator responses is parametrized by their spectral resolution $\delta / \Gamma_1$ and the measurement strength by the cooperativity $C_1 = C_2$.

\begin{figure}[tb!]
\includegraphics[width=3.375in]{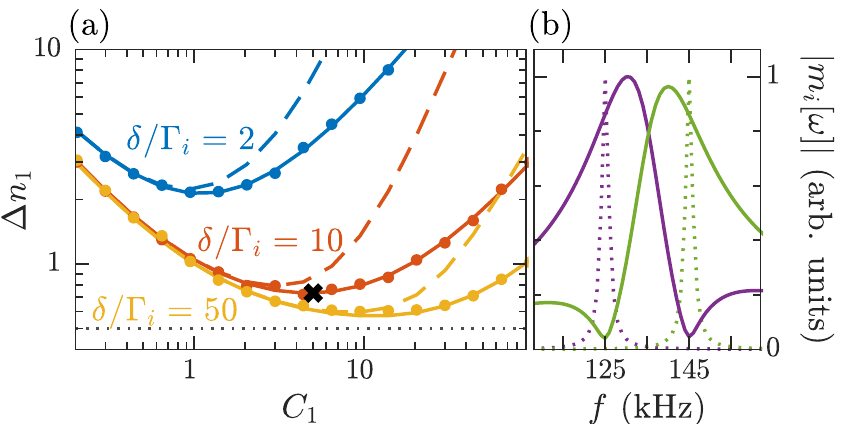}
\includegraphics[width=3.375in]{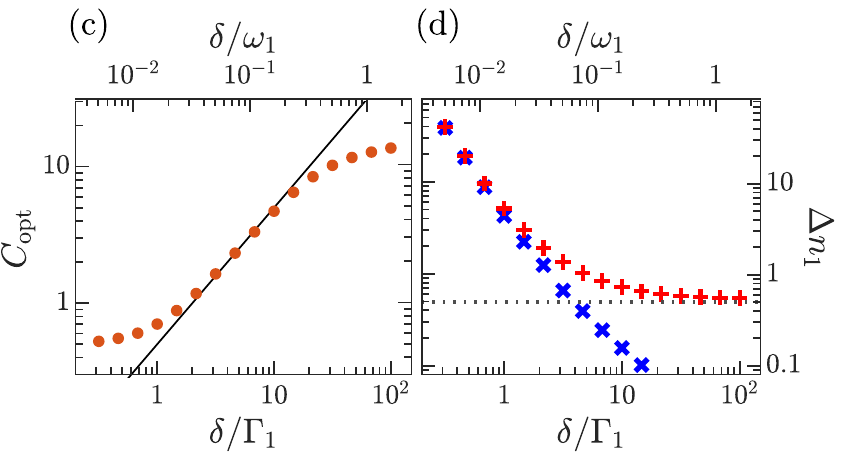}
\caption{
	\label{fig:twomode-sql}
	Optimal simultaneous two-mode retrodiction approaching the SQL.
	(a) Measurement imprecision $\imp_i$ for estimation of the state of one oscillator in a two oscillator system, using the optimal single-oscillator exponential filters (dashed lines) or multi-mode GLS filters (solid lines).
	The oscillator frequencies $\omega_1 / 2\pi = 125$ kHz and $\omega_2 = \omega_1 + \delta$ are detuned by $\delta = \{ 2, 10, 50 \} \Gamma_1$, with identical damping rates $\Gamma_1 = \Gamma_2 = 2\pi \times 2$ kHz to zero-temperature thermal baths.
	(b) Amplitude spectrum of one GLS filter for each oscillator, at the optimum measurement cooperativity (black cross in (a)).
	Motion of the other oscillator is suppressed in each by acquiring a notch at the position of the oscillator's response spectrum (dotted lines).
	(c) Numerically optimized measurement cooperativity that minimizes $\imp_1$ at a range of detunings $\delta$.
	For well-resolved oscillators, the optimal cooperativity is determined by $ \gFilt_\text{opt} = \delta $ (black line), where the corresponding single-oscillator filter linewidth equals the oscillator detuning.
	When $ \delta \gtrsim \omega_1 $, the cooperativity is again limited by the single-oscillator quality factor.
	(d) Minimized measurement imprecision $\imp_1$ (red pluses), corresponding to the optimal cooperativity in (c). State estimates for poorly-resolved oscillators $\delta / \Gamma_i \lesssim 1$ have strongly correlated errors $\langle | \Delta \est{a}_1^\dagger \Delta \est{a}_2 | \rangle$ (blue crosses).
}
\end{figure}

The total estimate uncertainty for the first oscillator $\imp_1$, defined by Eq.~\eqref{eq:osc-imprecision}, was numerically computed as a function of measurement cooperativity $C_1$ for a few different oscillator detunings, assuming perfect detection efficiency $\epsilon = 1$, with results shown in Fig.~\ref{fig:twomode-sql}a.
For sufficiently low-cooperativity measurements, the single-oscillator exponential filters defined by Eqs.~\eqref{eq:exp-filters} and \eqref{eq:optimal-filter-decay} achieve optimal results, with the measurement imprecision decreasing with increasing measurement strength $C_1$.
When the optimal single-oscillator filter decay rate $\gOpt$ approaches the oscillators' frequency difference $\delta$, the simple exponential filters fail to optimally distinguish the response of each oscillator, resulting in additional estimate imprecision.

The GLS filters defined by Eq.~\eqref{eq:gls-filters}, achieve reduced imprecision at high cooperativity.
These filters optimally distinguish the motion of each oscillator by acquiring a notch in the filter spectrum, shown in Fig.~\ref{fig:twomode-sql}b, that suppresses signal components in the frequency band of the other oscillator.
However, for larger cooperativity, the back-action broadened filters for each oscillator become increasingly indistinguishable.
The filter normalization matrix $\mat{J}$, defined by Eq.~\eqref{eq:normalization-matrix}, becomes nearly singular, resulting in growth of the normalized estimate imprecision.

The optimal measurement cooperativity for two-mode retrodiction at a given detuning $\delta$ was found by numerically minimizing the total estimate imprecision $\imp_i$, summarized in Fig.~\ref{fig:twomode-sql}c.
When $\Gamma_1 \ll \delta \ll \omega_1$, the optimal cooperativity is approximated by $\Copt = \delta / 2 \Gamma_1$, the threshold where the corresponding single-oscillator exponential filter linewidth defined by Eq.~\eqref{eq:diffusive-filter-decay} matches the oscillator detuning $\delta$.

At the optimal measurement cooperativity, the minimum estimate imprecision for each oscillator, shown in Fig.~\ref{fig:twomode-sql}d, approaches the SQL in the limit of well-resolved oscillators $\delta / \Gamma_i \gg 1$.
When the oscillators are not well resolved $\delta \lesssim \Gamma_1$, the filters cannot distinguish the motion of each oscillator.
The homodyne signal is always sensitive to the center of mass motion of the oscillators, but contains negligible information about their relative motion, resulting in strong correlated errors in the individual oscillator estimates, as displayed in Fig.~\ref{fig:twomode-sql}d, described by the off-diagonal blocks of the noise covariance matrices
\begin{align}
\langle \Delta \est{a}_1^\dagger \Delta \est{a}_2 \rangle = 
\frac{1}{2}
\begin{pmatrix}
1 & - i
\end{pmatrix}
\left[
\matTh_{12} + \matBA_{12} + \matSN_{12}
\right]
\begin{pmatrix}
1 \\ i
\end{pmatrix}
\text{.}
\end{align}
When $\delta \gtrsim \omega_i$, the optimal measurement cooperativity $\Copt$ is once more limited by the single-oscillator quality factor, as seen in Fig.~\ref{fig:single-oscillator}c.

\section{Experimental demonstration}
\label{sec:data}

\begin{figure}[tbh!]
\includegraphics[width=3.375in]{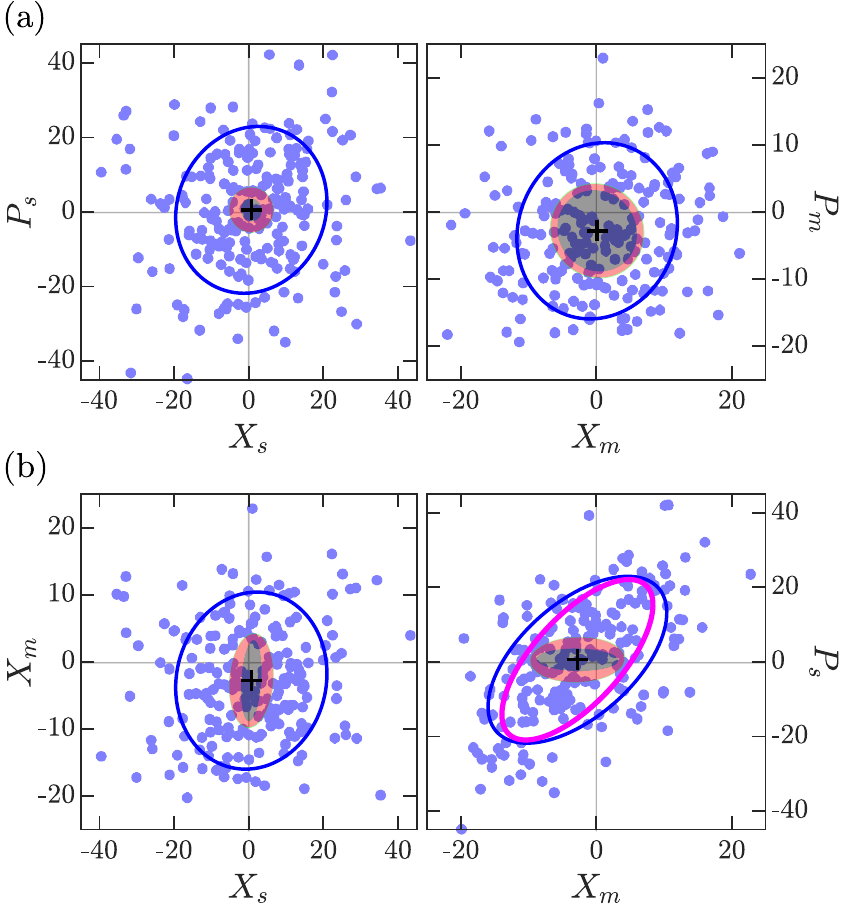}
\includegraphics[width=3.375in]{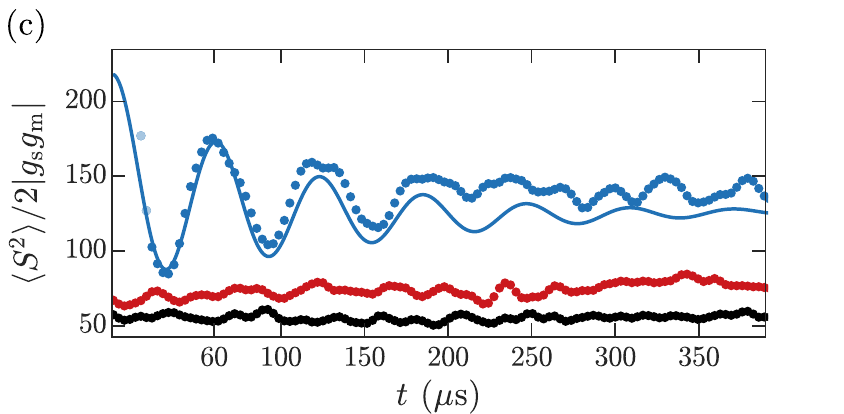}
\caption{
	\label{fig:data}
	Quadrature estimates from 215 repeated optodynamical measurements ($\nbar = 2.6$) of a correlated state of collective atomic spin and motion.
	(a) Individual state estimates of each mode indicate effective thermal states.
	The collective atomic spin (left), precessing near its highest energy state, represents a negative effective-mass oscillator, with independently calibrated parameters $\{g_s,\omega_s,\sigma_s,\Gamma_s\} = 2\pi \times \{18, 111, 0.2, 0\}$~kHz.
	The center-of-mass motion (right) provides a positive-mass oscillator, with independently calibrated parameters $\{g_s,\omega_s,\sigma_s,\Gamma_s\} = 2\pi \times \{27, 95, 0.4, 2.4\}$~kHz and $\nu_m = 2.7$.
	Estimate errors are dominated by thermal diffusion (red annulus) and measurement shot noise (black ellipse), with negligible measurement backaction (green annulus).
	(b)
	Cross correlation plots of in-phase (left) and out-of-phase (right) quadratures reveal a squeezed thermal state, with an imaginary correlation parameter characteristic of the negative-mass instability~\cite{Kohler2018}.
	Subtracting the systematic noise covariances gives the inferred state covariance (magenta ellipse).
	(c)
	The observed mean-squared signal (blue dots) agrees will with a reconstruction from the matched-filter covariance estimate (blue line).
	The transient beatnote from the initial correlations decays faster than the average signal relaxes to the equilibrium level (red dots), due to dephasing from inhomogeneous broadening.
}
\end{figure}

We previously employed this matched-filter analysis for retrodicting correlated states of a multi-mode optodynamical system, reported in Ref.~\cite{Kohler2018}.
In that work, correlations were generated through a negative-mass instability from collective coupling between the motion and spin of an atomic ensemble, resulting in resonant pair-creation analogous to a parametric amplifier.

Real-world measurements invariably bring additional complications, such as variations in system parameters and additional incoherent dynamics.
These must all be included in the preceding statistical analysis to accurately estimate the state covariance.
In particular, for Ref.~\cite{Kohler2018}, inhomogeneous broadening of the observed average homodyne PSD, due to shot-to-shot fluctuations of the oscillator frequencies $\omega_i$, caused dephasing of the ensemble-averaged signals and complicates the matched filter analysis.

To include this inhomogeneous broadening in the filter statistics, the oscillator frequency $\omega_i$ was treated as a classical random variable with variance $\sigma_i$.
The quadrature response functions $\vec{r}_i(t)$ are then also stochastic, with mean trajectory
\begin{align}
\langle \vec{ r }_i(t) \rangle = e^{-\Gamma_i t/2} e^{- \sigma_i^2 t^2 / 2} \begin{pmatrix} 
\cos \omega_i t \\
\sin \omega_i t
\end{pmatrix} \Theta(t)
\text{,}
\end{align}
which includes an additional decay envelope with dephasing rate $\sigma_i$, assuming a Gaussian frequency distribution.
The intrinsic damping rate $\Gamma_i$ and inhomogeneous dephasing rate $\sigma_i$ were experimentally calibrated by fitting the observed homodyne PSD with a Voigt profile.

This inhomogeneous broadening of the oscillator response motivates definition of modified quadrature filters
\begin{align}
\label{eq:avg-filters}
\vec{m}^\text{avg}_i(t) \equiv \langle \vec{r}_i(t) \rangle 
\text{,}
\end{align}
which are spectrally broadened to match the average PSD, in order to better capture the oscillator response across the distribution of frequencies.

The filter normalization matrix $\mat{J}$, defined in Eq.~\eqref{eq:normalization-matrix}, must also be considered as a matrix-valued random variable.
Assuming here that fluctuations of the system parameters are uncorrelated with the initial quadrature operators,
its expectation can be independently computed in Eq.~\eqref{eq:general-filter-mean} as
\begin{align}
\label{eq:stochastic-filter-mean} 
\langle \rawVec \rangle & = \langle \mat{J} \rangle \langle \hat{ \quadVec } \rangle
\text{.}
\end{align}
Individual quadrature estimates from data reported in Ref.~\cite{Kohler2018} are displayed in Fig.~\ref{fig:data}, obtained using filters defined by Eq.~\eqref{eq:avg-filters} and normalized by $\langle \mat{J} \rangle^{-1}$.

Estimation of the quadrature covariance is further complicated.
The 2nd-order moment of any pair of the unnormalized filter estimates, defined by Eq.~\eqref{eq:general-filter}, can be written
\begin{align}
\label{eq:stochastic-filter-2moment}
\langle \est{ q }_i \est{ q }_j \rangle = 
\sum_{kl} 
\langle \mat{J}_{ik} \mat{J}_{jl} \rangle \langle \hat{ \Q }_k \hat{ \Q }_l \rangle
+ \matTh'_{ij} + \matBA'_{ij} + \matSN'_{ij}
\text{.}
\end{align}
Here, the indices $i$, $j$, $k$, and $l$ run over all $2 \Nosc$ quadrature elements, unlike the block-matrix notation used above.
The covariances added from each noise source in Eq.~\eqref{eq:stochastic-filter-2moment} are given by the matrices
\begin{align}
\matSN' & = \Psn \, \int_{0}^{\tf}dt \,  \vec{m}(t) \vec{m}^{\Tr} (t) \text{,}
\\
\matTh' & = \sum_{k} \Bigl( \bathTemp_k + \frac{1}{2} \Bigr) \, \int_0^\tf d\tau \, \langle \mat{N}'_k(\tau) \mat{N}'^\Tr_k(\tau) \rangle \text{,}
\\
\matBA' & \approx \sum_{kl} \frac{ \sqrt{C_k C_l } }{2} \, \int_0^\tf d\tau \, \langle \mat{N}'_k(\tau) \mat{N}'^\Tr_l(\tau) \rangle \text{,}
\\
\mat{N}'_k(\tau) & = \sqrt{2 \Gamma_k} g_k \int_0^\tf dt \, \vec{m}(t) \vec{r}^\Tr_{k}(t - \tau) \text{.}
\end{align}
These expressions can be evaluated numerically as described in Appendix~\ref{app:response-correlations}, based on independent calibrations of the system parameters.
The system of $(2 \Nosc)^2$ equations defined by Eq.~\eqref{eq:stochastic-filter-2moment} can then be inverted to recover the 2nd-moments of the quadrature estimates and the added noise covariances, which define the covariance ellipses shown in Fig.~\ref{fig:data}a-b.

Finally, incoherent coupling in the experimental system between the spin and motion of individual atoms also resulted in additional thermal diffusion of the collective spin, which was included as an additional noise drive in Eq.~\eqref{eq:noise-drive}.
Diffusion from this interaction increased with $\nbar$, proportional to the measurement strength, and therefore limited the experimental system to low-cooperativity measurements, which prevented achieving retrodiction at the SQL.

\section{Conclusion}
\label{sec:conclusion}

In summary, we have derived a detailed model for linear state estimation from optical measurements of multi-mode optomechanical systems.
We have demonstrated that retrodiction of the past state of a single oscillator from high-cooperativity measurements approaches the SQL, when estimated with a matched filter that optimally suppresses the accumulated backaction noise.
Furthermore, we have described a general numerical method to derive optimal filters, which directly facilitates optimal estimation from simultaneous measurements of multiple oscillators.
We explored the optimal measurement strength for retrodiction of a multi-mode state and demonstrate additional constraints on the oscillator detuning in order to reach estimate imprecision at the SQL.
This work lays an experimentally motivated framework for simultaneous measurements of multi-mode systems, which provides an essential tool toward efficient measurements of many-body systems.

\appendix

\section{Convergence of sample covariance matrix}
\label{app:covariance-statistics}


Assuming the noise covariance matrices can be calculated to arbitrary precision, from independent calibration of the system and bath parameters, 
then the statistical uncertainty of the state covariance matrix inferred from Eq.~\eqref{eq:estimate-covariance} is determined solely by the uncertainty of the estimator covariance $\mat{\Sigma}$.

This covariance matrix can be estimated from an ensemble of $\Nsamp$ identical measurements $ \{ \est{ \quadVec }[i] \} $ according to the sample covariance
\begin{align}
\est{ \mat{ \Sigma } } = \frac{1}{\Nsamp - 1} \sum_i^{\Nsamp} \bigl( \est{ \quadVec }[i] - \est{ \vec{\mu} } \bigr) \bigl( \est{ \quadVec }[i] - \est{ \vec{\mu} } \bigr)^\Tr
\end{align}
where $\est{ \vec{ \mu } } = (1/\Nsamp) \sum_i \est{ \vec{ Q } }[i]$ is the sample mean.
The covariance estimator $\est{ \mat{ \Sigma}}$ is itself a random variable, which, assuming Gaussian initial states and input noise, is distributed according to the $2 \Nosc$-dimensional central Wishart distribution~\cite{Chatfield1980}
\begin{align}
(\Nsamp - 1) \est{ \mat{ \Sigma}} \sim W_{2 \Nosc} ( \mat{\Sigma}, \Nsamp - 1 )
\text{,}
\end{align}
with $(\Nsamp - 1)$ degrees of freedom.
 This distribution is the multi-dimensional generalization of the $\chi^2$ distribution, and allows calculation of estimate uncertainties from the variance of the matrix elements
\begin{align}
\operatorname{var} \bigl[ \est{ \mat{\Sigma} }_{ij} \bigr] = \frac{1}{\Nsamp - 1} ( \mat{ \Sigma }_{ij}^2 + \mat{ \Sigma }_{ii} \mat{ \Sigma }_{jj})
\text{.}
\end{align}
The uncertainty for estimating the multi-mode state covariance $\est{ \Sigma }$ from an ensemble of measurements, therefore, is reduced by minimizing the total systematic noise bias $\matTh + \matBA + \matSN$ added to the retrodicted estimate covariance $\mat{\Sigma}$ in Eq.~\eqref{eq:estimate-covariance}.

\section{Evaluation of estimate covariance from diffusive noise}
\label{app:response-correlations}

The covariance matrices of estimator noise added from thermal- and backaction-driven diffusion, defined in Eq.~\eqref{eq:thermal-covariance} and Eq.~\eqref{eq:backaction-covariance}, respectively, contain a triple integral over rapidly oscillatory integrands.
To aid in numerical evaluation, the integral which appears in each of these equations, describing the estimate covariance for correlated diffusion of oscillator $k$ and $l$ driven by a common bath, can be rewritten as
\begin{multline}
\label{eq:generic-covariance}
\int_0^\tf d\tau \, \mat{N}_k(\tau) \mat{N}^\Tr_l(\tau) =
\\
 g_k g_l \iint_0^\tf dt dt' \,  \mat{J}^{-1} \vec{m}(t) R_{kl}(t,t')
  \vec{m}^\Tr(t') [ \mat{J}^{-1}]^\Tr
\text{,}
\end{multline}
in terms of the generic oscillator two-time correlation for response to a common noise bath defined in Eq.~\eqref{eq:response-correlations}.

Generalizing to a complex response function 
\begin{align}
\rho_i(t) = e^{-(\Gamma_i/2 + i \omega_i) t} \Theta(t)
\text{,}
\end{align}
in terms of which
$ \vec{r}_i(t) = \begin{pmatrix}
\Re[ \rho_i(t) ], & - \Im[ \rho_i(t) ]
\end{pmatrix}^\Tr $
then Eq.~\eqref{eq:response-correlations} can be rewritten as
\begin{align}
R_{kl}(t,t') = \int_{0}^{\infty} d\tau \Re[ \rho^*_k(t - \tau) \rho_l(t' - \tau) ]
\text{.}
\end{align} 
This integral can be evaluated analytically, giving
\begin{align}
R_{kl}(t,t') = \Re \left[ \frac{ \rho_k^*(t - t') + \rho_l(t' - t) - \rho_k^*(t) \rho_l(t') }{ (\Gamma_k + \Gamma_l)/2 - i ( \omega_k - \omega_l ) } \right]
\text{,}
\end{align} 
 in terms of the complex oscillator responses.
This result can be used for efficient numerical evaluation of the double integral in Eq.~\eqref{eq:generic-covariance} for discretely sampled signals and filter functions.

When considering fluctuations of the oscillator frequencies in Sec.~\ref{sec:data}, it is necessary to evaluate the expectation of this two-time response product.
Assuming the variances of the oscillator frequencies are small compared to their difference, $\sigma_k + \sigma_l \ll |\omega_k - \omega_l|$ for $k \neq l$,
this expectation can be approximated by
\begin{align}
\langle R_{kl}(t,t') \rangle \approx \Re \left[ \frac{ \langle \rho_k^*(t - t') + \rho_l(t' - t) - \rho_k^*(t) \rho_l(t') \rangle }{ (\Gamma_k + \Gamma_l)/2 - i ( \omega_k - \omega_l ) } \right]
\text{.}
\end{align} 

\begin{acknowledgments}
The authors acknowledge helpful discussions with Mohan Sarovar, Zhishen Huang, and Klemmens Hammerer about related work on quantum state estimation, retrodiction, and smoothing.
Thanks also to Johannes Zeiher for feedback on the manuscript.
This work was supported by the Air Force Office of Scientific Research, through grants FA9550-14-1-0257 and FA9550-19-1-0328.
J.K. was supported by the U.S. Department of Defense through the National Defense Science and Engineering Graduate Fellowship program, and J.G. and E.D. by the National Science Foundation Graduate Fellowship.
\end{acknowledgments}

\bibliography{references,extra}

\begin{thebibliography}{53}%
\makeatletter
\providecommand \@ifxundefined [1]{%
 \@ifx{#1\undefined}
}%
\providecommand \@ifnum [1]{%
 \ifnum #1\expandafter \@firstoftwo
 \else \expandafter \@secondoftwo
 \fi
}%
\providecommand \@ifx [1]{%
 \ifx #1\expandafter \@firstoftwo
 \else \expandafter \@secondoftwo
 \fi
}%
\providecommand \natexlab [1]{#1}%
\providecommand \enquote  [1]{``#1''}%
\providecommand \bibnamefont  [1]{#1}%
\providecommand \bibfnamefont [1]{#1}%
\providecommand \citenamefont [1]{#1}%
\providecommand \href@noop [0]{\@secondoftwo}%
\providecommand \href [0]{\begingroup \@sanitize@url \@href}%
\providecommand \@href[1]{\@@startlink{#1}\@@href}%
\providecommand \@@href[1]{\endgroup#1\@@endlink}%
\providecommand \@sanitize@url [0]{\catcode `\\12\catcode `\$12\catcode
  `\&12\catcode `\#12\catcode `\^12\catcode `\_12\catcode `\%12\relax}%
\providecommand \@@startlink[1]{}%
\providecommand \@@endlink[0]{}%
\providecommand \url  [0]{\begingroup\@sanitize@url \@url }%
\providecommand \@url [1]{\endgroup\@href {#1}{\urlprefix }}%
\providecommand \urlprefix  [0]{URL }%
\providecommand \Eprint [0]{\href }%
\providecommand \doibase [0]{https://doi.org/}%
\providecommand \selectlanguage [0]{\@gobble}%
\providecommand \bibinfo  [0]{\@secondoftwo}%
\providecommand \bibfield  [0]{\@secondoftwo}%
\providecommand \translation [1]{[#1]}%
\providecommand \BibitemOpen [0]{}%
\providecommand \bibitemStop [0]{}%
\providecommand \bibitemNoStop [0]{.\EOS\space}%
\providecommand \EOS [0]{\spacefactor3000\relax}%
\providecommand \BibitemShut  [1]{\csname bibitem#1\endcsname}%
\let\auto@bib@innerbib\@empty
\bibitem [{\citenamefont {Jerger}\ \emph {et~al.}(2012)\citenamefont {Jerger},
  \citenamefont {Poletto}, \citenamefont {Macha}, \citenamefont {H{\"{u}}bner},
  \citenamefont {Il'ichev},\ and\ \citenamefont {Ustinov}}]{Jerger2012}%
  \BibitemOpen
  \bibfield  {author} {\bibinfo {author} {\bibfnamefont {M.}~\bibnamefont
  {Jerger}}, \bibinfo {author} {\bibfnamefont {S.}~\bibnamefont {Poletto}},
  \bibinfo {author} {\bibfnamefont {P.}~\bibnamefont {Macha}}, \bibinfo
  {author} {\bibfnamefont {U.}~\bibnamefont {H{\"{u}}bner}}, \bibinfo {author}
  {\bibfnamefont {E.}~\bibnamefont {Il'ichev}},\ and\ \bibinfo {author}
  {\bibfnamefont {A.~V.}\ \bibnamefont {Ustinov}},\ }\bibfield  {title}
  {\bibinfo {title} {Frequency division multiplexing readout and simultaneous
  manipulation of an array of flux qubits},\ }\href
  {https://doi.org/10.1063/1.4739454} {\bibfield  {journal} {\bibinfo
  {journal} {Appl. Phys. Lett.}\ }\textbf {\bibinfo {volume} {101}},\ \bibinfo
  {pages} {042604} (\bibinfo {year} {2012})}\BibitemShut {NoStop}%
\bibitem [{\citenamefont {Chen}\ \emph {et~al.}(2012)\citenamefont {Chen},
  \citenamefont {Sank}, \citenamefont {O'Malley}, \citenamefont {White},
  \citenamefont {Barends}, \citenamefont {Chiaro}, \citenamefont {Kelly},
  \citenamefont {Lucero}, \citenamefont {Mariantoni}, \citenamefont {Megrant},
  \citenamefont {Neill}, \citenamefont {Vainsencher}, \citenamefont {Wenner},
  \citenamefont {Yin}, \citenamefont {Cleland},\ and\ \citenamefont
  {Martinis}}]{Chen2012}%
  \BibitemOpen
  \bibfield  {author} {\bibinfo {author} {\bibfnamefont {Y.}~\bibnamefont
  {Chen}}, \bibinfo {author} {\bibfnamefont {D.}~\bibnamefont {Sank}}, \bibinfo
  {author} {\bibfnamefont {P.}~\bibnamefont {O'Malley}}, \bibinfo {author}
  {\bibfnamefont {T.}~\bibnamefont {White}}, \bibinfo {author} {\bibfnamefont
  {R.}~\bibnamefont {Barends}}, \bibinfo {author} {\bibfnamefont
  {B.}~\bibnamefont {Chiaro}}, \bibinfo {author} {\bibfnamefont
  {J.}~\bibnamefont {Kelly}}, \bibinfo {author} {\bibfnamefont
  {E.}~\bibnamefont {Lucero}}, \bibinfo {author} {\bibfnamefont
  {M.}~\bibnamefont {Mariantoni}}, \bibinfo {author} {\bibfnamefont
  {A.}~\bibnamefont {Megrant}}, \bibinfo {author} {\bibfnamefont
  {C.}~\bibnamefont {Neill}}, \bibinfo {author} {\bibfnamefont
  {A.}~\bibnamefont {Vainsencher}}, \bibinfo {author} {\bibfnamefont
  {J.}~\bibnamefont {Wenner}}, \bibinfo {author} {\bibfnamefont
  {Y.}~\bibnamefont {Yin}}, \bibinfo {author} {\bibfnamefont {A.~N.}\
  \bibnamefont {Cleland}},\ and\ \bibinfo {author} {\bibfnamefont {J.~M.}\
  \bibnamefont {Martinis}},\ }\bibfield  {title} {\bibinfo {title} {Multiplexed
  dispersive readout of superconducting phase qubits},\ }\href
  {https://doi.org/10.1063/1.4764940} {\bibfield  {journal} {\bibinfo
  {journal} {Appl. Phys. Lett.}\ }\textbf {\bibinfo {volume} {101}},\ \bibinfo
  {pages} {182601} (\bibinfo {year} {2012})}\BibitemShut {NoStop}%
\bibitem [{\citenamefont {Palomaki}\ \emph
  {et~al.}(2013{\natexlab{a}})\citenamefont {Palomaki}, \citenamefont {Teufel},
  \citenamefont {Simmonds},\ and\ \citenamefont {Lehnert}}]{Palomaki2013}%
  \BibitemOpen
  \bibfield  {author} {\bibinfo {author} {\bibfnamefont {T.~A.}\ \bibnamefont
  {Palomaki}}, \bibinfo {author} {\bibfnamefont {J.~D.}\ \bibnamefont
  {Teufel}}, \bibinfo {author} {\bibfnamefont {R.~W.}\ \bibnamefont
  {Simmonds}},\ and\ \bibinfo {author} {\bibfnamefont {K.~W.}\ \bibnamefont
  {Lehnert}},\ }\bibfield  {title} {\bibinfo {title} {Entangling mechanical
  motion with microwave fields},\ }\href
  {https://doi.org/10.1126/science.1244563} {\bibfield  {journal} {\bibinfo
  {journal} {Science}\ }\textbf {\bibinfo {volume} {342}},\ \bibinfo {pages}
  {710} (\bibinfo {year} {2013}{\natexlab{a}})}\BibitemShut {NoStop}%
\bibitem [{\citenamefont {Aspelmeyer}\ \emph {et~al.}(2014)\citenamefont
  {Aspelmeyer}, \citenamefont {Kippenberg},\ and\ \citenamefont
  {Marquardt}}]{Aspelmeyer2014}%
  \BibitemOpen
  \bibfield  {author} {\bibinfo {author} {\bibfnamefont {M.}~\bibnamefont
  {Aspelmeyer}}, \bibinfo {author} {\bibfnamefont {T.~J.}\ \bibnamefont
  {Kippenberg}},\ and\ \bibinfo {author} {\bibfnamefont {F.}~\bibnamefont
  {Marquardt}},\ }\bibfield  {title} {\bibinfo {title} {Cavity optomechanics},\
  }\href {https://doi.org/10.1103/RevModPhys.86.1391} {\bibfield  {journal}
  {\bibinfo  {journal} {Rev. Mod. Phys.}\ }\textbf {\bibinfo {volume} {86}},\
  \bibinfo {pages} {1391} (\bibinfo {year} {2014})}\BibitemShut {NoStop}%
\bibitem [{\citenamefont {Leroux}\ \emph {et~al.}(2010)\citenamefont {Leroux},
  \citenamefont {Schleier-Smith},\ and\ \citenamefont
  {Vuleti{\'{c}}}}]{Leroux2010}%
  \BibitemOpen
  \bibfield  {author} {\bibinfo {author} {\bibfnamefont {I.~D.}\ \bibnamefont
  {Leroux}}, \bibinfo {author} {\bibfnamefont {M.~H.}\ \bibnamefont
  {Schleier-Smith}},\ and\ \bibinfo {author} {\bibfnamefont {V.}~\bibnamefont
  {Vuleti{\'{c}}}},\ }\bibfield  {title} {\bibinfo {title} {Implementation of
  cavity squeezing of a collective atomic spin},\ }\href
  {https://doi.org/10.1103/PhysRevLett.104.073602} {\bibfield  {journal}
  {\bibinfo  {journal} {Phys. Rev. Lett.}\ }\textbf {\bibinfo {volume} {104}},\
  \bibinfo {pages} {073602} (\bibinfo {year} {2010})}\BibitemShut {NoStop}%
\bibitem [{\citenamefont {Vasilakis}\ \emph {et~al.}(2015)\citenamefont
  {Vasilakis}, \citenamefont {Shen}, \citenamefont {Jensen}, \citenamefont
  {Balabas}, \citenamefont {Salart}, \citenamefont {Chen},\ and\ \citenamefont
  {Polzik}}]{Vasilakis2015}%
  \BibitemOpen
  \bibfield  {author} {\bibinfo {author} {\bibfnamefont {G.}~\bibnamefont
  {Vasilakis}}, \bibinfo {author} {\bibfnamefont {H.}~\bibnamefont {Shen}},
  \bibinfo {author} {\bibfnamefont {K.}~\bibnamefont {Jensen}}, \bibinfo
  {author} {\bibfnamefont {M.}~\bibnamefont {Balabas}}, \bibinfo {author}
  {\bibfnamefont {D.}~\bibnamefont {Salart}}, \bibinfo {author} {\bibfnamefont
  {B.}~\bibnamefont {Chen}},\ and\ \bibinfo {author} {\bibfnamefont {E.~S.}\
  \bibnamefont {Polzik}},\ }\bibfield  {title} {\bibinfo {title} {Generation of
  a squeezed state of an oscillator by stroboscopic back-action-evading
  measurement},\ }\href {https://doi.org/10.1038/nphys3280} {\bibfield
  {journal} {\bibinfo  {journal} {Nat. Phys.}\ }\textbf {\bibinfo {volume}
  {11}},\ \bibinfo {pages} {389} (\bibinfo {year} {2015})}\BibitemShut
  {NoStop}%
\bibitem [{\citenamefont {Kohler}\ \emph {et~al.}(2017)\citenamefont {Kohler},
  \citenamefont {Spethmann}, \citenamefont {Schreppler},\ and\ \citenamefont
  {Stamper-Kurn}}]{Kohler2017}%
  \BibitemOpen
  \bibfield  {author} {\bibinfo {author} {\bibfnamefont {J.}~\bibnamefont
  {Kohler}}, \bibinfo {author} {\bibfnamefont {N.}~\bibnamefont {Spethmann}},
  \bibinfo {author} {\bibfnamefont {S.}~\bibnamefont {Schreppler}},\ and\
  \bibinfo {author} {\bibfnamefont {D.~M.}\ \bibnamefont {Stamper-Kurn}},\
  }\bibfield  {title} {\bibinfo {title} {Cavity-assisted measurement and
  coherent control of collective atomic spin oscillators},\ }\href
  {https://doi.org/10.1103/PhysRevLett.118.063604} {\bibfield  {journal}
  {\bibinfo  {journal} {Phys. Rev. Lett.}\ }\textbf {\bibinfo {volume} {118}},\
  \bibinfo {pages} {063604} (\bibinfo {year} {2017})}\BibitemShut {NoStop}%
\bibitem [{\citenamefont {Zhang}\ \emph {et~al.}(2016)\citenamefont {Zhang},
  \citenamefont {Zhu}, \citenamefont {Zou},\ and\ \citenamefont
  {Tang}}]{Zhang2015a}%
  \BibitemOpen
  \bibfield  {author} {\bibinfo {author} {\bibfnamefont {X.}~\bibnamefont
  {Zhang}}, \bibinfo {author} {\bibfnamefont {N.}~\bibnamefont {Zhu}}, \bibinfo
  {author} {\bibfnamefont {C.-L.}\ \bibnamefont {Zou}},\ and\ \bibinfo {author}
  {\bibfnamefont {H.~X.}\ \bibnamefont {Tang}},\ }\bibfield  {title} {\bibinfo
  {title} {Optomagnonic whispering gallery microresonators},\ }\href
  {https://doi.org/10.1103/PhysRevLett.117.123605} {\bibfield  {journal}
  {\bibinfo  {journal} {Phys. Rev. Lett.}\ }\textbf {\bibinfo {volume} {117}},\
  \bibinfo {pages} {123605} (\bibinfo {year} {2016})}\BibitemShut {NoStop}%
\bibitem [{\citenamefont {Haigh}\ \emph {et~al.}(2015)\citenamefont {Haigh},
  \citenamefont {Langenfeld}, \citenamefont {Lambert}, \citenamefont
  {Baumberg}, \citenamefont {Ramsay}, \citenamefont {Nunnenkamp},\ and\
  \citenamefont {Ferguson}}]{Haigh2015}%
  \BibitemOpen
  \bibfield  {author} {\bibinfo {author} {\bibfnamefont {J.~A.}\ \bibnamefont
  {Haigh}}, \bibinfo {author} {\bibfnamefont {S.}~\bibnamefont {Langenfeld}},
  \bibinfo {author} {\bibfnamefont {N.~J.}\ \bibnamefont {Lambert}}, \bibinfo
  {author} {\bibfnamefont {J.~J.}\ \bibnamefont {Baumberg}}, \bibinfo {author}
  {\bibfnamefont {A.~J.}\ \bibnamefont {Ramsay}}, \bibinfo {author}
  {\bibfnamefont {A.}~\bibnamefont {Nunnenkamp}},\ and\ \bibinfo {author}
  {\bibfnamefont {A.~J.}\ \bibnamefont {Ferguson}},\ }\bibfield  {title}
  {\bibinfo {title} {Magneto-optical coupling in whispering-gallery-mode
  resonators},\ }\href {https://doi.org/10.1103/PhysRevA.92.063845} {\bibfield
  {journal} {\bibinfo  {journal} {Phys. Rev. A}\ }\textbf {\bibinfo {volume}
  {92}},\ \bibinfo {pages} {063845} (\bibinfo {year} {2015})}\BibitemShut
  {NoStop}%
\bibitem [{\citenamefont {Osada}\ \emph {et~al.}(2016)\citenamefont {Osada},
  \citenamefont {Hisatomi}, \citenamefont {Noguchi}, \citenamefont {Tabuchi},
  \citenamefont {Yamazaki}, \citenamefont {Usami}, \citenamefont {Sadgrove},
  \citenamefont {Yalla}, \citenamefont {Nomura},\ and\ \citenamefont
  {Nakamura}}]{Osada2016}%
  \BibitemOpen
  \bibfield  {author} {\bibinfo {author} {\bibfnamefont {A.}~\bibnamefont
  {Osada}}, \bibinfo {author} {\bibfnamefont {R.}~\bibnamefont {Hisatomi}},
  \bibinfo {author} {\bibfnamefont {A.}~\bibnamefont {Noguchi}}, \bibinfo
  {author} {\bibfnamefont {Y.}~\bibnamefont {Tabuchi}}, \bibinfo {author}
  {\bibfnamefont {R.}~\bibnamefont {Yamazaki}}, \bibinfo {author}
  {\bibfnamefont {K.}~\bibnamefont {Usami}}, \bibinfo {author} {\bibfnamefont
  {M.}~\bibnamefont {Sadgrove}}, \bibinfo {author} {\bibfnamefont
  {R.}~\bibnamefont {Yalla}}, \bibinfo {author} {\bibfnamefont
  {M.}~\bibnamefont {Nomura}},\ and\ \bibinfo {author} {\bibfnamefont
  {Y.}~\bibnamefont {Nakamura}},\ }\bibfield  {title} {\bibinfo {title} {Cavity
  optomagnonics with spin-orbit coupled photons},\ }\href
  {https://doi.org/10.1103/PhysRevLett.116.223601} {\bibfield  {journal}
  {\bibinfo  {journal} {Phys. Rev. Lett.}\ }\textbf {\bibinfo {volume} {116}},\
  \bibinfo {pages} {223601} (\bibinfo {year} {2016})}\BibitemShut {NoStop}%
\bibitem [{\citenamefont {Liu}\ \emph {et~al.}(2016)\citenamefont {Liu},
  \citenamefont {Zhang}, \citenamefont {Tang},\ and\ \citenamefont
  {Flatt{\'{e}}}}]{Liu2016}%
  \BibitemOpen
  \bibfield  {author} {\bibinfo {author} {\bibfnamefont {T.}~\bibnamefont
  {Liu}}, \bibinfo {author} {\bibfnamefont {X.}~\bibnamefont {Zhang}}, \bibinfo
  {author} {\bibfnamefont {H.~X.}\ \bibnamefont {Tang}},\ and\ \bibinfo
  {author} {\bibfnamefont {M.~E.}\ \bibnamefont {Flatt{\'{e}}}},\ }\bibfield
  {title} {\bibinfo {title} {Optomagnonics in magnetic solids},\ }\href
  {https://doi.org/10.1103/PhysRevB.94.060405} {\bibfield  {journal} {\bibinfo
  {journal} {Phys. Rev. B}\ }\textbf {\bibinfo {volume} {94}},\ \bibinfo
  {pages} {060405(R)} (\bibinfo {year} {2016})}\BibitemShut {NoStop}%
\bibitem [{\citenamefont {{Viola Kusminskiy}}\ \emph
  {et~al.}(2016)\citenamefont {{Viola Kusminskiy}}, \citenamefont {Tang},\ and\
  \citenamefont {Marquardt}}]{Kusminskiy2016}%
  \BibitemOpen
  \bibfield  {author} {\bibinfo {author} {\bibfnamefont {S.}~\bibnamefont
  {{Viola Kusminskiy}}}, \bibinfo {author} {\bibfnamefont {H.~X.}\ \bibnamefont
  {Tang}},\ and\ \bibinfo {author} {\bibfnamefont {F.}~\bibnamefont
  {Marquardt}},\ }\bibfield  {title} {\bibinfo {title} {Coupled spin-light
  dynamics in cavity optomagnonics},\ }\href
  {https://doi.org/10.1103/PhysRevA.94.033821} {\bibfield  {journal} {\bibinfo
  {journal} {Phys. Rev. A}\ }\textbf {\bibinfo {volume} {94}},\ \bibinfo
  {pages} {033821} (\bibinfo {year} {2016})}\BibitemShut {NoStop}%
\bibitem [{\citenamefont {Botter}\ \emph {et~al.}(2013)\citenamefont {Botter},
  \citenamefont {Brooks}, \citenamefont {Schreppler}, \citenamefont {Brahms},\
  and\ \citenamefont {Stamper-Kurn}}]{Botter2013}%
  \BibitemOpen
  \bibfield  {author} {\bibinfo {author} {\bibfnamefont {T.}~\bibnamefont
  {Botter}}, \bibinfo {author} {\bibfnamefont {D.~W.~C.}\ \bibnamefont
  {Brooks}}, \bibinfo {author} {\bibfnamefont {S.}~\bibnamefont {Schreppler}},
  \bibinfo {author} {\bibfnamefont {N.}~\bibnamefont {Brahms}},\ and\ \bibinfo
  {author} {\bibfnamefont {D.~M.}\ \bibnamefont {Stamper-Kurn}},\ }\bibfield
  {title} {\bibinfo {title} {Optical readout of the quantum collective motion
  of an array of atomic ensembles},\ }\href
  {https://doi.org/10.1103/PhysRevLett.110.153001} {\bibfield  {journal}
  {\bibinfo  {journal} {Phys. Rev. Lett.}\ }\textbf {\bibinfo {volume} {110}},\
  \bibinfo {pages} {153001} (\bibinfo {year} {2013})}\BibitemShut {NoStop}%
\bibitem [{\citenamefont {Shkarin}\ \emph {et~al.}(2014)\citenamefont
  {Shkarin}, \citenamefont {Flowers-Jacobs}, \citenamefont {Hoch},
  \citenamefont {Kashkanova}, \citenamefont {Deutsch}, \citenamefont
  {Reichel},\ and\ \citenamefont {Harris}}]{Shkarin2014}%
  \BibitemOpen
  \bibfield  {author} {\bibinfo {author} {\bibfnamefont {A.~B.}\ \bibnamefont
  {Shkarin}}, \bibinfo {author} {\bibfnamefont {N.~E.}\ \bibnamefont
  {Flowers-Jacobs}}, \bibinfo {author} {\bibfnamefont {S.~W.}\ \bibnamefont
  {Hoch}}, \bibinfo {author} {\bibfnamefont {A.~D.}\ \bibnamefont
  {Kashkanova}}, \bibinfo {author} {\bibfnamefont {C.}~\bibnamefont {Deutsch}},
  \bibinfo {author} {\bibfnamefont {J.}~\bibnamefont {Reichel}},\ and\ \bibinfo
  {author} {\bibfnamefont {J.~G.~E.}\ \bibnamefont {Harris}},\ }\bibfield
  {title} {\bibinfo {title} {Optically mediated hybridization between two
  mechanical modes},\ }\href {https://doi.org/10.1103/PhysRevLett.112.013602}
  {\bibfield  {journal} {\bibinfo  {journal} {Phys. Rev. Lett.}\ }\textbf
  {\bibinfo {volume} {112}},\ \bibinfo {pages} {013602} (\bibinfo {year}
  {2014})}\BibitemShut {NoStop}%
\bibitem [{\citenamefont {Spethmann}\ \emph {et~al.}(2016)\citenamefont
  {Spethmann}, \citenamefont {Kohler}, \citenamefont {Schreppler},
  \citenamefont {Buchmann},\ and\ \citenamefont
  {Stamper-Kurn}}]{Spethmann2015}%
  \BibitemOpen
  \bibfield  {author} {\bibinfo {author} {\bibfnamefont {N.}~\bibnamefont
  {Spethmann}}, \bibinfo {author} {\bibfnamefont {J.}~\bibnamefont {Kohler}},
  \bibinfo {author} {\bibfnamefont {S.}~\bibnamefont {Schreppler}}, \bibinfo
  {author} {\bibfnamefont {L.}~\bibnamefont {Buchmann}},\ and\ \bibinfo
  {author} {\bibfnamefont {D.~M.}\ \bibnamefont {Stamper-Kurn}},\ }\bibfield
  {title} {\bibinfo {title} {Cavity-mediated coupling of mechanical oscillators
  limited by quantum back-action},\ }\href {https://doi.org/10.1038/nphys3515}
  {\bibfield  {journal} {\bibinfo  {journal} {Nat. Phys.}\ }\textbf {\bibinfo
  {volume} {12}},\ \bibinfo {pages} {27} (\bibinfo {year} {2016})}\BibitemShut
  {NoStop}%
\bibitem [{\citenamefont {Kohler}\ \emph {et~al.}(2018)\citenamefont {Kohler},
  \citenamefont {Gerber}, \citenamefont {Dowd},\ and\ \citenamefont
  {Stamper-Kurn}}]{Kohler2018}%
  \BibitemOpen
  \bibfield  {author} {\bibinfo {author} {\bibfnamefont {J.}~\bibnamefont
  {Kohler}}, \bibinfo {author} {\bibfnamefont {J.~A.}\ \bibnamefont {Gerber}},
  \bibinfo {author} {\bibfnamefont {E.}~\bibnamefont {Dowd}},\ and\ \bibinfo
  {author} {\bibfnamefont {D.~M.}\ \bibnamefont {Stamper-Kurn}},\ }\bibfield
  {title} {\bibinfo {title} {Negative-mass instability of the spin and motion
  of an atomic gas driven by optical cavity backaction},\ }\href
  {https://doi.org/10.1103/PhysRevLett.120.013601} {\bibfield  {journal}
  {\bibinfo  {journal} {Phys. Rev. Lett.}\ }\textbf {\bibinfo {volume} {120}},\
  \bibinfo {pages} {013601} (\bibinfo {year} {2018})}\BibitemShut {NoStop}%
\bibitem [{\citenamefont {Gammelmark}\ \emph {et~al.}(2013)\citenamefont
  {Gammelmark}, \citenamefont {Julsgaard},\ and\ \citenamefont
  {M{\o}lmer}}]{Gammelmark2013}%
  \BibitemOpen
  \bibfield  {author} {\bibinfo {author} {\bibfnamefont {S.}~\bibnamefont
  {Gammelmark}}, \bibinfo {author} {\bibfnamefont {B.}~\bibnamefont
  {Julsgaard}},\ and\ \bibinfo {author} {\bibfnamefont {K.}~\bibnamefont
  {M{\o}lmer}},\ }\bibfield  {title} {\bibinfo {title} {Past quantum states of
  a monitored system},\ }\href {https://doi.org/10.1103/PhysRevLett.111.160401}
  {\bibfield  {journal} {\bibinfo  {journal} {Phys. Rev. Lett.}\ }\textbf
  {\bibinfo {volume} {111}},\ \bibinfo {pages} {160401} (\bibinfo {year}
  {2013})}\BibitemShut {NoStop}%
\bibitem [{\citenamefont {Tsang}(2009)}]{Tsang2009b}%
  \BibitemOpen
  \bibfield  {author} {\bibinfo {author} {\bibfnamefont {M.}~\bibnamefont
  {Tsang}},\ }\bibfield  {title} {\bibinfo {title} {Time-symmetric quantum
  theory of smoothing},\ }\href
  {https://doi.org/10.1103/PhysRevLett.102.250403} {\bibfield  {journal}
  {\bibinfo  {journal} {Phys. Rev. Lett.}\ }\textbf {\bibinfo {volume} {102}},\
  \bibinfo {pages} {250403} (\bibinfo {year} {2009})}\BibitemShut {NoStop}%
\bibitem [{\citenamefont {Zhang}\ and\ \citenamefont
  {M{\o}lmer}(2017)}]{Zhang2017}%
  \BibitemOpen
  \bibfield  {author} {\bibinfo {author} {\bibfnamefont {J.}~\bibnamefont
  {Zhang}}\ and\ \bibinfo {author} {\bibfnamefont {K.}~\bibnamefont
  {M{\o}lmer}},\ }\bibfield  {title} {\bibinfo {title} {Prediction and
  retrodiction with continuously monitored {G}aussian states},\ }\href
  {https://doi.org/10.1103/PhysRevA.96.062131} {\bibfield  {journal} {\bibinfo
  {journal} {Phys. Rev. A}\ }\textbf {\bibinfo {volume} {96}},\ \bibinfo
  {pages} {062131} (\bibinfo {year} {2017})}\BibitemShut {NoStop}%
\bibitem [{\citenamefont {Huang}\ and\ \citenamefont
  {Sarovar}(2018)}]{Huang2018}%
  \BibitemOpen
  \bibfield  {author} {\bibinfo {author} {\bibfnamefont {Z.}~\bibnamefont
  {Huang}}\ and\ \bibinfo {author} {\bibfnamefont {M.}~\bibnamefont
  {Sarovar}},\ }\bibfield  {title} {\bibinfo {title} {Smoothing of {G}aussian
  quantum dynamics for force detection},\ }\href
  {https://doi.org/10.1103/PhysRevA.97.042106} {\bibfield  {journal} {\bibinfo
  {journal} {Phys. Rev. A}\ }\textbf {\bibinfo {volume} {97}},\ \bibinfo
  {pages} {042106} (\bibinfo {year} {2018})}\BibitemShut {NoStop}%
\bibitem [{\citenamefont {Lammers}(2018)}]{LammersThesis}%
  \BibitemOpen
  \bibfield  {author} {\bibinfo {author} {\bibfnamefont {J.}~\bibnamefont
  {Lammers}},\ }\emph {\bibinfo {title} {State preparation and verification in
  continuously measured quantum systems}},\ \href
  {https://doi.org/10.15488/3854} {Ph.D. thesis},\ \bibinfo  {school}
  {Gottfried Wilhelm Leibniz Universit{\"{a}}t} (\bibinfo {year}
  {2018})\BibitemShut {NoStop}%
\bibitem [{\citenamefont {Lammers}\ and\ \citenamefont
  {Hammerer}()}]{Lammers2018}%
  \BibitemOpen
  \bibfield  {author} {\bibinfo {author} {\bibfnamefont {J.}~\bibnamefont
  {Lammers}}\ and\ \bibinfo {author} {\bibfnamefont {K.}~\bibnamefont
  {Hammerer}},\ }\bibinfo {note} {to be published}\BibitemShut {NoStop}%
\bibitem [{\citenamefont {Rossi}\ \emph {et~al.}(2019)\citenamefont {Rossi},
  \citenamefont {Mason}, \citenamefont {Chen},\ and\ \citenamefont
  {Schliesser}}]{Rossi2019}%
  \BibitemOpen
  \bibfield  {author} {\bibinfo {author} {\bibfnamefont {M.}~\bibnamefont
  {Rossi}}, \bibinfo {author} {\bibfnamefont {D.}~\bibnamefont {Mason}},
  \bibinfo {author} {\bibfnamefont {J.}~\bibnamefont {Chen}},\ and\ \bibinfo
  {author} {\bibfnamefont {A.}~\bibnamefont {Schliesser}},\ }\bibfield  {title}
  {\bibinfo {title} {{Observing and Verifying the Quantum Trajectory of a
  Mechanical Resonator}},\ }\href
  {https://doi.org/10.1103/PhysRevLett.123.163601} {\bibfield  {journal}
  {\bibinfo  {journal} {Phys. Rev. Lett.}\ }\textbf {\bibinfo {volume} {123}},\
  \bibinfo {pages} {163601} (\bibinfo {year} {2019})}\BibitemShut {NoStop}%
\bibitem [{\citenamefont {M{\"{u}}ller-Ebhardt}\ \emph
  {et~al.}(2009)\citenamefont {M{\"{u}}ller-Ebhardt}, \citenamefont {Rehbein},
  \citenamefont {Li}, \citenamefont {Mino}, \citenamefont {Somiya},
  \citenamefont {Schnabel}, \citenamefont {Danzmann},\ and\ \citenamefont
  {Chen}}]{Muller-Ebhardt2009}%
  \BibitemOpen
  \bibfield  {author} {\bibinfo {author} {\bibfnamefont {H.}~\bibnamefont
  {M{\"{u}}ller-Ebhardt}}, \bibinfo {author} {\bibfnamefont {H.}~\bibnamefont
  {Rehbein}}, \bibinfo {author} {\bibfnamefont {C.}~\bibnamefont {Li}},
  \bibinfo {author} {\bibfnamefont {Y.}~\bibnamefont {Mino}}, \bibinfo {author}
  {\bibfnamefont {K.}~\bibnamefont {Somiya}}, \bibinfo {author} {\bibfnamefont
  {R.}~\bibnamefont {Schnabel}}, \bibinfo {author} {\bibfnamefont
  {K.}~\bibnamefont {Danzmann}},\ and\ \bibinfo {author} {\bibfnamefont
  {Y.}~\bibnamefont {Chen}},\ }\bibfield  {title} {\bibinfo {title}
  {Quantum-state preparation and macroscopic entanglement in gravitational-wave
  detectors},\ }\href {https://doi.org/10.1103/PhysRevA.80.043802} {\bibfield
  {journal} {\bibinfo  {journal} {Phys. Rev. A}\ }\textbf {\bibinfo {volume}
  {80}},\ \bibinfo {pages} {043802} (\bibinfo {year} {2009})}\BibitemShut
  {NoStop}%
\bibitem [{\citenamefont {Miao}\ \emph {et~al.}(2010)\citenamefont {Miao},
  \citenamefont {Danilishin}, \citenamefont {M{\"{u}}ller-Ebhardt},
  \citenamefont {Rehbein}, \citenamefont {Somiya},\ and\ \citenamefont
  {Chen}}]{Miao2010}%
  \BibitemOpen
  \bibfield  {author} {\bibinfo {author} {\bibfnamefont {H.}~\bibnamefont
  {Miao}}, \bibinfo {author} {\bibfnamefont {S.}~\bibnamefont {Danilishin}},
  \bibinfo {author} {\bibfnamefont {H.}~\bibnamefont {M{\"{u}}ller-Ebhardt}},
  \bibinfo {author} {\bibfnamefont {H.}~\bibnamefont {Rehbein}}, \bibinfo
  {author} {\bibfnamefont {K.}~\bibnamefont {Somiya}},\ and\ \bibinfo {author}
  {\bibfnamefont {Y.}~\bibnamefont {Chen}},\ }\bibfield  {title} {\bibinfo
  {title} {Probing macroscopic quantum states with a sub-{H}eisenberg
  accuracy},\ }\href {https://doi.org/10.1103/PhysRevA.81.012114} {\bibfield
  {journal} {\bibinfo  {journal} {Phys. Rev. A}\ }\textbf {\bibinfo {volume}
  {81}},\ \bibinfo {pages} {012114} (\bibinfo {year} {2010})}\BibitemShut
  {NoStop}%
\bibitem [{\citenamefont {Moore}\ \emph {et~al.}(2016)\citenamefont {Moore},
  \citenamefont {Tufarelli}, \citenamefont {Paternostro},\ and\ \citenamefont
  {Ferraro}}]{Moore2016}%
  \BibitemOpen
  \bibfield  {author} {\bibinfo {author} {\bibfnamefont {D.~W.}\ \bibnamefont
  {Moore}}, \bibinfo {author} {\bibfnamefont {T.}~\bibnamefont {Tufarelli}},
  \bibinfo {author} {\bibfnamefont {M.}~\bibnamefont {Paternostro}},\ and\
  \bibinfo {author} {\bibfnamefont {A.}~\bibnamefont {Ferraro}},\ }\bibfield
  {title} {\bibinfo {title} {Quantum state reconstruction of an oscillator
  network in an optomechanical setting},\ }\href
  {https://doi.org/10.1103/PhysRevA.94.053811} {\bibfield  {journal} {\bibinfo
  {journal} {Phys. Rev. A}\ }\textbf {\bibinfo {volume} {94}},\ \bibinfo
  {pages} {053811} (\bibinfo {year} {2016})}\BibitemShut {NoStop}%
\bibitem [{\citenamefont {Wainstein}\ and\ \citenamefont
  {Zubakov}(1970)}]{Wainstein1970}%
  \BibitemOpen
  \bibfield  {author} {\bibinfo {author} {\bibfnamefont {L.~A.}\ \bibnamefont
  {Wainstein}}\ and\ \bibinfo {author} {\bibfnamefont {V.~D.}\ \bibnamefont
  {Zubakov}},\ }\href@noop {} {\emph {\bibinfo {title} {Extraction of Signals
  from Noise}}}\ (\bibinfo  {publisher} {Dover Publ., Incorporated},\ \bibinfo
  {address} {New York},\ \bibinfo {year} {1970})\BibitemShut {NoStop}%
\bibitem [{\citenamefont {Abbott}\ \emph {et~al.}(2004)\citenamefont {Abbott}
  \emph {et~al.}}]{LIGOAnalysis}%
  \BibitemOpen
  \bibfield  {author} {\bibinfo {author} {\bibfnamefont {B.~P.}\ \bibnamefont
  {Abbott}} \emph {et~al.} (\bibinfo {collaboration} {{LIGO} Scientific}),\
  }\bibfield  {title} {\bibinfo {title} {Analysis of {LIGO} data for
  gravitational waves from binary neutron stars},\ }\href
  {https://doi.org/10.1103/PhysRevD.69.122001} {\bibfield  {journal} {\bibinfo
  {journal} {Phys. Rev. D}\ }\textbf {\bibinfo {volume} {69}},\ \bibinfo
  {pages} {122001} (\bibinfo {year} {2004})}\BibitemShut {NoStop}%
\bibitem [{\citenamefont {Caves}(1980)}]{Caves1980a}%
  \BibitemOpen
  \bibfield  {author} {\bibinfo {author} {\bibfnamefont {C.~M.}\ \bibnamefont
  {Caves}},\ }\bibfield  {title} {\bibinfo {title} {Quantum-mechanical
  radiation-pressure fluctuations in an interferometer},\ }\href
  {https://doi.org/10.1103/PhysRevLett.45.75} {\bibfield  {journal} {\bibinfo
  {journal} {Phys. Rev. Lett.}\ }\textbf {\bibinfo {volume} {45}},\ \bibinfo
  {pages} {75} (\bibinfo {year} {1980})}\BibitemShut {NoStop}%
\bibitem [{\citenamefont {Braginsky}\ \emph {et~al.}(1992)\citenamefont
  {Braginsky}, \citenamefont {Khalili},\ and\ \citenamefont
  {Thorne}}]{BKT_QuantumNoise_1992}%
  \BibitemOpen
  \bibfield  {author} {\bibinfo {author} {\bibfnamefont {V.~B.}\ \bibnamefont
  {Braginsky}}, \bibinfo {author} {\bibfnamefont {F.~Y.}\ \bibnamefont
  {Khalili}},\ and\ \bibinfo {author} {\bibfnamefont {K.~S.}\ \bibnamefont
  {Thorne}},\ }\href {https://doi.org/10.1017/CBO9780511622748} {\emph
  {\bibinfo {title} {Quantum Measurement}}}\ (\bibinfo  {publisher} {Cambridge
  University Press},\ \bibinfo {year} {1992})\BibitemShut {NoStop}%
\bibitem [{\citenamefont {Nha}\ \emph {et~al.}(2010)\citenamefont {Nha},
  \citenamefont {Milburn},\ and\ \citenamefont {Carmichael}}]{Nha2010}%
  \BibitemOpen
  \bibfield  {author} {\bibinfo {author} {\bibfnamefont {H.}~\bibnamefont
  {Nha}}, \bibinfo {author} {\bibfnamefont {G.~J.}\ \bibnamefont {Milburn}},\
  and\ \bibinfo {author} {\bibfnamefont {H.~J.}\ \bibnamefont {Carmichael}},\
  }\bibfield  {title} {\bibinfo {title} {Linear amplification and quantum
  cloning for non-{G}aussian continuous variables},\ }\href
  {https://doi.org/10.1088/1367-2630/12/10/103010} {\bibfield  {journal}
  {\bibinfo  {journal} {New J. Phys.}\ }\textbf {\bibinfo {volume} {12}},\
  \bibinfo {pages} {103010} (\bibinfo {year} {2010})}\BibitemShut {NoStop}%
\bibitem [{\citenamefont {Lei}\ \emph {et~al.}(2016)\citenamefont {Lei},
  \citenamefont {Weinstein}, \citenamefont {Suh}, \citenamefont {Wollman},
  \citenamefont {Kronwald}, \citenamefont {Marquardt}, \citenamefont {Clerk},\
  and\ \citenamefont {Schwab}}]{Lei2016}%
  \BibitemOpen
  \bibfield  {author} {\bibinfo {author} {\bibfnamefont {C.~U.}\ \bibnamefont
  {Lei}}, \bibinfo {author} {\bibfnamefont {A.~J.}\ \bibnamefont {Weinstein}},
  \bibinfo {author} {\bibfnamefont {J.}~\bibnamefont {Suh}}, \bibinfo {author}
  {\bibfnamefont {E.~E.}\ \bibnamefont {Wollman}}, \bibinfo {author}
  {\bibfnamefont {A.}~\bibnamefont {Kronwald}}, \bibinfo {author}
  {\bibfnamefont {F.}~\bibnamefont {Marquardt}}, \bibinfo {author}
  {\bibfnamefont {A.~A.}\ \bibnamefont {Clerk}},\ and\ \bibinfo {author}
  {\bibfnamefont {K.~C.}\ \bibnamefont {Schwab}},\ }\bibfield  {title}
  {\bibinfo {title} {Quantum nondemolition measurement of a quantum squeezed
  state beyond the 3 {dB} limit},\ }\href
  {https://doi.org/10.1103/PhysRevLett.117.100801} {\bibfield  {journal}
  {\bibinfo  {journal} {Phys. Rev. Lett.}\ }\textbf {\bibinfo {volume} {117}},\
  \bibinfo {pages} {100801} (\bibinfo {year} {2016})}\BibitemShut {NoStop}%
\bibitem [{\citenamefont {Ockeloen-Korppi}\ \emph {et~al.}(2016)\citenamefont
  {Ockeloen-Korppi}, \citenamefont {Damsk{\"{a}}gg}, \citenamefont
  {Pirkkalainen}, \citenamefont {Clerk}, \citenamefont {Woolley},\ and\
  \citenamefont {Sillanp{\"{a}}{\"{a}}}}]{Ockeloen-Korppi2016}%
  \BibitemOpen
  \bibfield  {author} {\bibinfo {author} {\bibfnamefont {C.~F.}\ \bibnamefont
  {Ockeloen-Korppi}}, \bibinfo {author} {\bibfnamefont {E.}~\bibnamefont
  {Damsk{\"{a}}gg}}, \bibinfo {author} {\bibfnamefont {J.-M.}\ \bibnamefont
  {Pirkkalainen}}, \bibinfo {author} {\bibfnamefont {A.~A.}\ \bibnamefont
  {Clerk}}, \bibinfo {author} {\bibfnamefont {M.~J.}\ \bibnamefont {Woolley}},\
  and\ \bibinfo {author} {\bibfnamefont {M.~A.}\ \bibnamefont
  {Sillanp{\"{a}}{\"{a}}}},\ }\bibfield  {title} {\bibinfo {title} {Quantum
  backaction evading measurement of collective mechanical modes},\ }\href
  {https://doi.org/10.1103/PhysRevLett.117.140401} {\bibfield  {journal}
  {\bibinfo  {journal} {Phys. Rev. Lett.}\ }\textbf {\bibinfo {volume} {117}},\
  \bibinfo {pages} {140401} (\bibinfo {year} {2016})}\BibitemShut {NoStop}%
\bibitem [{\citenamefont {M{\o}ller}\ \emph {et~al.}(2017)\citenamefont
  {M{\o}ller}, \citenamefont {Thomas}, \citenamefont {Vasilakis}, \citenamefont
  {Zeuthen}, \citenamefont {Tsaturyan}, \citenamefont {Balabas}, \citenamefont
  {Jensen}, \citenamefont {Schliesser}, \citenamefont {Hammerer},\ and\
  \citenamefont {Polzik}}]{Moller2017}%
  \BibitemOpen
  \bibfield  {author} {\bibinfo {author} {\bibfnamefont {C.~B.}\ \bibnamefont
  {M{\o}ller}}, \bibinfo {author} {\bibfnamefont {R.~A.}\ \bibnamefont
  {Thomas}}, \bibinfo {author} {\bibfnamefont {G.}~\bibnamefont {Vasilakis}},
  \bibinfo {author} {\bibfnamefont {E.}~\bibnamefont {Zeuthen}}, \bibinfo
  {author} {\bibfnamefont {Y.}~\bibnamefont {Tsaturyan}}, \bibinfo {author}
  {\bibfnamefont {M.}~\bibnamefont {Balabas}}, \bibinfo {author} {\bibfnamefont
  {K.}~\bibnamefont {Jensen}}, \bibinfo {author} {\bibfnamefont
  {A.}~\bibnamefont {Schliesser}}, \bibinfo {author} {\bibfnamefont
  {K.}~\bibnamefont {Hammerer}},\ and\ \bibinfo {author} {\bibfnamefont
  {E.~S.}\ \bibnamefont {Polzik}},\ }\bibfield  {title} {\bibinfo {title}
  {Quantum back-action-evading measurement of motion in a negative mass
  reference frame},\ }\href {https://doi.org/10.1038/nature22980} {\bibfield
  {journal} {\bibinfo  {journal} {Nature}\ }\textbf {\bibinfo {volume} {547}},\
  \bibinfo {pages} {191} (\bibinfo {year} {2017})}\BibitemShut {NoStop}%
\bibitem [{\citenamefont {Sheard}\ \emph {et~al.}(2004)\citenamefont {Sheard},
  \citenamefont {Gray}, \citenamefont {Mow-Lowry}, \citenamefont {McClelland},\
  and\ \citenamefont {Whitcomb}}]{Sheard2004}%
  \BibitemOpen
  \bibfield  {author} {\bibinfo {author} {\bibfnamefont {B.~S.}\ \bibnamefont
  {Sheard}}, \bibinfo {author} {\bibfnamefont {M.~B.}\ \bibnamefont {Gray}},
  \bibinfo {author} {\bibfnamefont {C.~M.}\ \bibnamefont {Mow-Lowry}}, \bibinfo
  {author} {\bibfnamefont {D.~E.}\ \bibnamefont {McClelland}},\ and\ \bibinfo
  {author} {\bibfnamefont {S.~E.}\ \bibnamefont {Whitcomb}},\ }\bibfield
  {title} {\bibinfo {title} {Observation and characterization of an optical
  spring},\ }\href {https://doi.org/10.1103/PhysRevA.69.051801} {\bibfield
  {journal} {\bibinfo  {journal} {Phys. Rev. A}\ }\textbf {\bibinfo {volume}
  {69}},\ \bibinfo {pages} {051801(R)} (\bibinfo {year} {2004})}\BibitemShut
  {NoStop}%
\bibitem [{\citenamefont {Corbitt}\ \emph {et~al.}(2006)\citenamefont
  {Corbitt}, \citenamefont {Ottaway}, \citenamefont {Innerhofer}, \citenamefont
  {Pelc},\ and\ \citenamefont {Mavalvala}}]{Corbitt2006}%
  \BibitemOpen
  \bibfield  {author} {\bibinfo {author} {\bibfnamefont {T.}~\bibnamefont
  {Corbitt}}, \bibinfo {author} {\bibfnamefont {D.}~\bibnamefont {Ottaway}},
  \bibinfo {author} {\bibfnamefont {E.}~\bibnamefont {Innerhofer}}, \bibinfo
  {author} {\bibfnamefont {J.}~\bibnamefont {Pelc}},\ and\ \bibinfo {author}
  {\bibfnamefont {N.}~\bibnamefont {Mavalvala}},\ }\bibfield  {title} {\bibinfo
  {title} {Measurement of radiation-pressure-induced optomechanical dynamics in
  a suspended {F}abry-{P}erot cavity},\ }\href
  {https://doi.org/10.1103/PhysRevA.74.021802} {\bibfield  {journal} {\bibinfo
  {journal} {Phys. Rev. A}\ }\textbf {\bibinfo {volume} {74}},\ \bibinfo
  {pages} {021802(R)} (\bibinfo {year} {2006})}\BibitemShut {NoStop}%
\bibitem [{\citenamefont {Arcizet}\ \emph {et~al.}(2006)\citenamefont
  {Arcizet}, \citenamefont {Cohadon}, \citenamefont {Briant}, \citenamefont
  {Pinard},\ and\ \citenamefont {Heidmann}}]{Arcizet2006}%
  \BibitemOpen
  \bibfield  {author} {\bibinfo {author} {\bibfnamefont {O.}~\bibnamefont
  {Arcizet}}, \bibinfo {author} {\bibfnamefont {P.-F.}\ \bibnamefont
  {Cohadon}}, \bibinfo {author} {\bibfnamefont {T.}~\bibnamefont {Briant}},
  \bibinfo {author} {\bibfnamefont {M.}~\bibnamefont {Pinard}},\ and\ \bibinfo
  {author} {\bibfnamefont {A.}~\bibnamefont {Heidmann}},\ }\bibfield  {title}
  {\bibinfo {title} {Radiation-pressure cooling and optomechanical instability
  of a micromirror},\ }\href {https://doi.org/10.1038/nature05244} {\bibfield
  {journal} {\bibinfo  {journal} {Nature (London)}\ }\textbf {\bibinfo {volume}
  {444}},\ \bibinfo {pages} {71} (\bibinfo {year} {2006})}\BibitemShut
  {NoStop}%
\bibitem [{\citenamefont {Gigan}\ \emph {et~al.}(2006)\citenamefont {Gigan},
  \citenamefont {B{\"{o}}hm}, \citenamefont {Paternostro}, \citenamefont
  {Blaser}, \citenamefont {Langer}, \citenamefont {Hertzberg}, \citenamefont
  {Schwab}, \citenamefont {B{\"{a}}uerle}, \citenamefont {Aspelmeyer},\ and\
  \citenamefont {Zeilinger}}]{Gigan2006}%
  \BibitemOpen
  \bibfield  {author} {\bibinfo {author} {\bibfnamefont {S.}~\bibnamefont
  {Gigan}}, \bibinfo {author} {\bibfnamefont {H.~R.}\ \bibnamefont
  {B{\"{o}}hm}}, \bibinfo {author} {\bibfnamefont {M.}~\bibnamefont
  {Paternostro}}, \bibinfo {author} {\bibfnamefont {F.}~\bibnamefont {Blaser}},
  \bibinfo {author} {\bibfnamefont {G.}~\bibnamefont {Langer}}, \bibinfo
  {author} {\bibfnamefont {J.~B.}\ \bibnamefont {Hertzberg}}, \bibinfo {author}
  {\bibfnamefont {K.~C.}\ \bibnamefont {Schwab}}, \bibinfo {author}
  {\bibfnamefont {D.}~\bibnamefont {B{\"{a}}uerle}}, \bibinfo {author}
  {\bibfnamefont {M.}~\bibnamefont {Aspelmeyer}},\ and\ \bibinfo {author}
  {\bibfnamefont {A.}~\bibnamefont {Zeilinger}},\ }\bibfield  {title} {\bibinfo
  {title} {Self-cooling of a micromirror by radiation pressure},\ }\href
  {https://doi.org/10.1038/nature05273} {\bibfield  {journal} {\bibinfo
  {journal} {Nature (London)}\ }\textbf {\bibinfo {volume} {444}},\ \bibinfo
  {pages} {67} (\bibinfo {year} {2006})}\BibitemShut {NoStop}%
\bibitem [{\citenamefont {Schliesser}\ \emph {et~al.}(2006)\citenamefont
  {Schliesser}, \citenamefont {Del'Haye}, \citenamefont {Nooshi}, \citenamefont
  {Vahala},\ and\ \citenamefont {Kippenberg}}]{Schliesser2006}%
  \BibitemOpen
  \bibfield  {author} {\bibinfo {author} {\bibfnamefont {A.}~\bibnamefont
  {Schliesser}}, \bibinfo {author} {\bibfnamefont {P.}~\bibnamefont
  {Del'Haye}}, \bibinfo {author} {\bibfnamefont {N.}~\bibnamefont {Nooshi}},
  \bibinfo {author} {\bibfnamefont {K.~J.}\ \bibnamefont {Vahala}},\ and\
  \bibinfo {author} {\bibfnamefont {T.~J.}\ \bibnamefont {Kippenberg}},\
  }\bibfield  {title} {\bibinfo {title} {Radiation pressure cooling of a
  micromechanical oscillator using dynamical backaction},\ }\href
  {https://doi.org/10.1103/PhysRevLett.97.243905} {\bibfield  {journal}
  {\bibinfo  {journal} {Phys. Rev. Lett.}\ }\textbf {\bibinfo {volume} {97}},\
  \bibinfo {pages} {243905} (\bibinfo {year} {2006})}\BibitemShut {NoStop}%
\bibitem [{\citenamefont {Gardiner}\ and\ \citenamefont
  {Collett}(1985)}]{Gardiner1985}%
  \BibitemOpen
  \bibfield  {author} {\bibinfo {author} {\bibfnamefont {C.~W.}\ \bibnamefont
  {Gardiner}}\ and\ \bibinfo {author} {\bibfnamefont {M.~J.}\ \bibnamefont
  {Collett}},\ }\bibfield  {title} {\bibinfo {title} {Input and output in
  damped quantum systems: Quantum stochastic differential equations and the
  master equation},\ }\href {https://doi.org/10.1103/PhysRevA.31.3761}
  {\bibfield  {journal} {\bibinfo  {journal} {Phys. Rev. A}\ }\textbf {\bibinfo
  {volume} {31}},\ \bibinfo {pages} {3761} (\bibinfo {year}
  {1985})}\BibitemShut {NoStop}%
\bibitem [{Note1()}]{Note1}%
  \BibitemOpen
  \bibinfo {note} {Assuming $g_i > 0$, without loss of generality.}\BibitemShut
  {Stop}%
\bibitem [{\citenamefont {McElroy}(1967)}]{McElroy1967}%
  \BibitemOpen
  \bibfield  {author} {\bibinfo {author} {\bibfnamefont {F.~W.}\ \bibnamefont
  {McElroy}},\ }\bibfield  {title} {\bibinfo {title} {A necessary and
  sufficient condition that ordinary least-squares estimators be best linear
  unbiased},\ }\href {https://doi.org/10.2307/2283779} {\bibfield  {journal}
  {\bibinfo  {journal} {J. Am. Stat. Assoc.}\ }\textbf {\bibinfo {volume}
  {62}},\ \bibinfo {pages} {1302} (\bibinfo {year} {1967})}\BibitemShut
  {NoStop}%
\bibitem [{\citenamefont {Aitken}(1936)}]{Aitken1936}%
  \BibitemOpen
  \bibfield  {author} {\bibinfo {author} {\bibfnamefont {A.~C.}\ \bibnamefont
  {Aitken}},\ }\bibfield  {title} {\bibinfo {title} {{IV}.--{O}n least squares
  and linear combination of observations.},\ }\href
  {https://doi.org/10.1017/S0370164600014346} {\bibfield  {journal} {\bibinfo
  {journal} {Proc. R. Soc. Edinburgh}\ }\textbf {\bibinfo {volume} {55}},\
  \bibinfo {pages} {42} (\bibinfo {year} {1936})}\BibitemShut {NoStop}%
\bibitem [{\citenamefont {Walls}\ and\ \citenamefont
  {Milburn}(2008)}]{WallsAndMilburn}%
  \BibitemOpen
  \bibfield  {author} {\bibinfo {author} {\bibfnamefont {D.~F.}\ \bibnamefont
  {Walls}}\ and\ \bibinfo {author} {\bibfnamefont {G.~J.}\ \bibnamefont
  {Milburn}},\ }\href {https://doi.org/10.1007/978-3-540-28574-8} {\emph
  {\bibinfo {title} {Quantum Optics}}},\ \bibinfo {edition} {2nd}\ ed.\
  (\bibinfo  {publisher} {Springer-Verlag Berlin Heidelberg},\ \bibinfo
  {address} {Berlin},\ \bibinfo {year} {2008})\BibitemShut {NoStop}%
\bibitem [{\citenamefont {Rao}(1976)}]{Rao1976}%
  \BibitemOpen
  \bibfield  {author} {\bibinfo {author} {\bibfnamefont {C.~R.}\ \bibnamefont
  {Rao}},\ }\bibfield  {title} {\bibinfo {title} {Estimation of parameters in a
  linear model},\ }\href {https://www.jstor.org/stable/2958576} {\bibfield
  {journal} {\bibinfo  {journal} {Ann. Stat.}\ }\textbf {\bibinfo {volume}
  {4}},\ \bibinfo {pages} {1023} (\bibinfo {year} {1976})}\BibitemShut
  {NoStop}%
\bibitem [{\citenamefont {Teufel}\ \emph {et~al.}(2009)\citenamefont {Teufel},
  \citenamefont {Donner}, \citenamefont {Castellanos-Beltran}, \citenamefont
  {Harlow},\ and\ \citenamefont {Lehnert}}]{Teufel2009}%
  \BibitemOpen
  \bibfield  {author} {\bibinfo {author} {\bibfnamefont {J.~D.}\ \bibnamefont
  {Teufel}}, \bibinfo {author} {\bibfnamefont {T.}~\bibnamefont {Donner}},
  \bibinfo {author} {\bibfnamefont {M.~A.}\ \bibnamefont
  {Castellanos-Beltran}}, \bibinfo {author} {\bibfnamefont {J.~W.}\
  \bibnamefont {Harlow}},\ and\ \bibinfo {author} {\bibfnamefont {K.~W.}\
  \bibnamefont {Lehnert}},\ }\bibfield  {title} {\bibinfo {title}
  {Nanomechanical motion measured with an imprecision below that at the
  standard quantum limit},\ }\href {https://doi.org/10.1038/nnano.2009.343}
  {\bibfield  {journal} {\bibinfo  {journal} {Nat. Nanotechnol.}\ }\textbf
  {\bibinfo {volume} {4}},\ \bibinfo {pages} {820} (\bibinfo {year}
  {2009})}\BibitemShut {NoStop}%
\bibitem [{\citenamefont {Anetsberger}\ \emph {et~al.}(2010)\citenamefont
  {Anetsberger}, \citenamefont {Gavartin}, \citenamefont {Arcizet},
  \citenamefont {Unterreithmeier}, \citenamefont {Weig}, \citenamefont
  {Gorodetsky}, \citenamefont {Kotthaus},\ and\ \citenamefont
  {Kippenberg}}]{Anetsberger2010}%
  \BibitemOpen
  \bibfield  {author} {\bibinfo {author} {\bibfnamefont {G.}~\bibnamefont
  {Anetsberger}}, \bibinfo {author} {\bibfnamefont {E.}~\bibnamefont
  {Gavartin}}, \bibinfo {author} {\bibfnamefont {O.}~\bibnamefont {Arcizet}},
  \bibinfo {author} {\bibfnamefont {Q.~P.}\ \bibnamefont {Unterreithmeier}},
  \bibinfo {author} {\bibfnamefont {E.~M.}\ \bibnamefont {Weig}}, \bibinfo
  {author} {\bibfnamefont {M.~L.}\ \bibnamefont {Gorodetsky}}, \bibinfo
  {author} {\bibfnamefont {J.~P.}\ \bibnamefont {Kotthaus}},\ and\ \bibinfo
  {author} {\bibfnamefont {T.~J.}\ \bibnamefont {Kippenberg}},\ }\bibfield
  {title} {\bibinfo {title} {Measuring nanomechanical motion with an
  imprecision below the standard quantum limit},\ }\href
  {https://doi.org/10.1103/PhysRevA.82.061804} {\bibfield  {journal} {\bibinfo
  {journal} {Phys. Rev. A}\ }\textbf {\bibinfo {volume} {82}},\ \bibinfo
  {pages} {061804(R)} (\bibinfo {year} {2010})}\BibitemShut {NoStop}%
\bibitem [{\citenamefont {Rossi}\ \emph {et~al.}(2018)\citenamefont {Rossi},
  \citenamefont {Mason}, \citenamefont {Chen}, \citenamefont {Tsaturyan},\ and\
  \citenamefont {Schliesser}}]{Rossi2018}%
  \BibitemOpen
  \bibfield  {author} {\bibinfo {author} {\bibfnamefont {M.}~\bibnamefont
  {Rossi}}, \bibinfo {author} {\bibfnamefont {D.}~\bibnamefont {Mason}},
  \bibinfo {author} {\bibfnamefont {J.}~\bibnamefont {Chen}}, \bibinfo {author}
  {\bibfnamefont {Y.}~\bibnamefont {Tsaturyan}},\ and\ \bibinfo {author}
  {\bibfnamefont {A.}~\bibnamefont {Schliesser}},\ }\bibfield  {title}
  {\bibinfo {title} {Measurement-based quantum control of mechanical motion},\
  }\href {https://doi.org/10.1038/s41586-018-0643-8} {\bibfield  {journal}
  {\bibinfo  {journal} {Nature}\ }\textbf {\bibinfo {volume} {563}},\ \bibinfo
  {pages} {53} (\bibinfo {year} {2018})}\BibitemShut {NoStop}%
\bibitem [{\citenamefont {Schreppler}\ \emph {et~al.}(2014)\citenamefont
  {Schreppler}, \citenamefont {Spethmann}, \citenamefont {Brahms},
  \citenamefont {Botter}, \citenamefont {Barrios},\ and\ \citenamefont
  {Stamper-Kurn}}]{Schreppler2014}%
  \BibitemOpen
  \bibfield  {author} {\bibinfo {author} {\bibfnamefont {S.}~\bibnamefont
  {Schreppler}}, \bibinfo {author} {\bibfnamefont {N.}~\bibnamefont
  {Spethmann}}, \bibinfo {author} {\bibfnamefont {N.}~\bibnamefont {Brahms}},
  \bibinfo {author} {\bibfnamefont {T.}~\bibnamefont {Botter}}, \bibinfo
  {author} {\bibfnamefont {M.}~\bibnamefont {Barrios}},\ and\ \bibinfo {author}
  {\bibfnamefont {D.~M.}\ \bibnamefont {Stamper-Kurn}},\ }\bibfield  {title}
  {\bibinfo {title} {Optically measuring force near the standard quantum
  limit},\ }\href {https://doi.org/10.1126/science.1249850} {\bibfield
  {journal} {\bibinfo  {journal} {Science}\ }\textbf {\bibinfo {volume}
  {344}},\ \bibinfo {pages} {1486} (\bibinfo {year} {2014})}\BibitemShut
  {NoStop}%
\bibitem [{\citenamefont {Mason}\ \emph {et~al.}(2019)\citenamefont {Mason},
  \citenamefont {Chen}, \citenamefont {Rossi}, \citenamefont {Tsaturyan},\ and\
  \citenamefont {Schliesser}}]{Mason2018}%
  \BibitemOpen
  \bibfield  {author} {\bibinfo {author} {\bibfnamefont {D.}~\bibnamefont
  {Mason}}, \bibinfo {author} {\bibfnamefont {J.}~\bibnamefont {Chen}},
  \bibinfo {author} {\bibfnamefont {M.}~\bibnamefont {Rossi}}, \bibinfo
  {author} {\bibfnamefont {Y.}~\bibnamefont {Tsaturyan}},\ and\ \bibinfo
  {author} {\bibfnamefont {A.}~\bibnamefont {Schliesser}},\ }\bibfield  {title}
  {\bibinfo {title} {{Continuous force and displacement measurement below the
  standard quantum limit}},\ }\href {https://doi.org/10.1038/s41567-019-0533-5}
  {\bibfield  {journal} {\bibinfo  {journal} {Nat. Phys.}\ }\textbf {\bibinfo
  {volume} {15}},\ \bibinfo {pages} {745} (\bibinfo {year} {2019})}\BibitemShut
  {NoStop}%
\bibitem [{\citenamefont {Abbott}\ \emph {et~al.}(2016)\citenamefont {Abbott}
  \emph {et~al.}}]{LIGODetection}%
  \BibitemOpen
  \bibfield  {author} {\bibinfo {author} {\bibfnamefont {B.~P.}\ \bibnamefont
  {Abbott}} \emph {et~al.} (\bibinfo {collaboration} {LIGO Scientific
  Collaboration and Virgo Collaboration}),\ }\bibfield  {title} {\bibinfo
  {title} {Observation of gravitational waves from a binary black hole
  merger},\ }\href {https://doi.org/10.1103/PhysRevLett.116.061102} {\bibfield
  {journal} {\bibinfo  {journal} {Phys. Rev. Lett.}\ }\textbf {\bibinfo
  {volume} {116}},\ \bibinfo {pages} {061102} (\bibinfo {year}
  {2016})}\BibitemShut {NoStop}%
\bibitem [{\citenamefont {Palomaki}\ \emph
  {et~al.}(2013{\natexlab{b}})\citenamefont {Palomaki}, \citenamefont {Harlow},
  \citenamefont {Teufel}, \citenamefont {Simmonds},\ and\ \citenamefont
  {Lehnert}}]{Palomaki2013a}%
  \BibitemOpen
  \bibfield  {author} {\bibinfo {author} {\bibfnamefont {T.~A.}\ \bibnamefont
  {Palomaki}}, \bibinfo {author} {\bibfnamefont {J.~W.}\ \bibnamefont
  {Harlow}}, \bibinfo {author} {\bibfnamefont {J.~D.}\ \bibnamefont {Teufel}},
  \bibinfo {author} {\bibfnamefont {R.~W.}\ \bibnamefont {Simmonds}},\ and\
  \bibinfo {author} {\bibfnamefont {K.~W.}\ \bibnamefont {Lehnert}},\
  }\bibfield  {title} {\bibinfo {title} {Coherent state transfer between
  itinerant microwave fields and a mechanical oscillator},\ }\href
  {https://doi.org/10.1038/nature11915} {\bibfield  {journal} {\bibinfo
  {journal} {Nature}\ }\textbf {\bibinfo {volume} {495}},\ \bibinfo {pages}
  {210} (\bibinfo {year} {2013}{\natexlab{b}})}\BibitemShut {NoStop}%
\bibitem [{\citenamefont {Chatfield}\ and\ \citenamefont
  {Collins}(1980)}]{Chatfield1980}%
  \BibitemOpen
  \bibfield  {author} {\bibinfo {author} {\bibfnamefont {C.}~\bibnamefont
  {Chatfield}}\ and\ \bibinfo {author} {\bibfnamefont {A.~J.}\ \bibnamefont
  {Collins}},\ }\href {https://doi.org/10.1007/978-1-4899-3184-9} {\emph
  {\bibinfo {title} {Introduction to Multivariate Analysis}}}\ (\bibinfo
  {publisher} {Springer US},\ \bibinfo {address} {Boston, MA},\ \bibinfo {year}
  {1980})\BibitemShut {NoStop}%
\end{thebibliography}%

\end{document}